# First, do NOHARM: a medical safety benchmark and randomized study of physician and AI teaming on clinical consultations

David Wu[1,2,3*], Fateme Nateghi Haredasht[4*], Saloni Kumar Maharaj[5], Priyank Jain[6,3], Jessica Tran[5], Matthew Gwiazdon[7], Arjun Rustagi[8], Jenelle Jindal[9], Jacob M. Koshy[10,3], Vinay Kadiyala[10,3], Anup Agarwal[6,3], Bassman Tappuni[11,3], Brianna French[12], Sirus Jesudasen[13], Christopher V. Cosgriff[14,15,3,16], Rebanta Chakraborty[17,3], Jillian Caldwell[18], Susan Ziolkowski[18], David J. Iberri[19], Robert Diep[19], Rahul S. Dalal[20,3], Kira L. Newman[21], Kristin Galetta[9], J. Carl Pallais[22,3], Nancy Wei[23,3], Kathleen M. Buchheit[24], David I. Hong[24], Vartan Pahalyants[25], Ernest Y. Lee[26,27], Allen Shih[28,3], Tamara B. Kaplan[29,3], Vishnu Ravi[18], Sarita Khemani[5], Thomas A. Buckley[30], April S. Liang[5], Daniel Shirvani[31], Advait Patil[3], Nicholas Marshall[32], Kanav Chopra[4], Joel Koh[33], Adi Badhwar[4], Anastasia Perez[4,34], Austin J. Schoeffler[35], Mahbuba Tusty[36], Chase M. Walton[10], Liam G. McCoy[37,10,38], David J. H. Wu[39], Yingjie Weng[40], Sumant Ranji[41], Kevin Schulman[34], Nigam H. Shah[4], Jason Hom[5], Arnold Milstein[34], Arjun K. Manrai[30], Adam Rodman[10,3†], Jonathan H. Chen[4,34,5†], and Ethan Goh[4,34†]

[1]Harvard Combined Dermatology Program, Boston, MA, USA

[2]Department of Dermatology, Mass General Brigham, Boston, MA, USA

[3]Harvard Medical School, Boston, MA, USA

[4]Stanford Division of Computational Medicine, Stanford University, Stanford, CA, USA

[5]Division of Hospital Medicine, Department of Medicine, Stanford University School of Medicine, Stanford, CA, USA

[6]Department of Medicine, Cambridge Health Alliance, Cambridge, MA, USA

[7]Beth Israel Deaconess Hospital–Plymouth, Plymouth, MA, USA

[8]Department of Medicine, University of California, San Francisco, San Francisco, CA, USA

[9]Department of Neurology, Stanford University School of Medicine, Stanford, CA, USA

[10]Department of Medicine, Beth Israel Deaconess Medical Center, Boston, MA, USA

[11]Division of Cardiology, Department of Medicine, Cambridge Health Alliance, Cambridge, MA, USA

[12]Department of Cardiovascular Medicine, Summa Health System, Akron, OH, USA

[13]Division of Allergy, Pulmonary, and Critical Care Medicine, Department of Medicine, University of Wisconsin-Madison, Madison, WI, USA

[14]Division of Pulmonary and Critical Care Medicine, Department of Medicine, Massachusetts General Hospital, Boston, MA, USA

[15]Center for Immunology and Inflammatory Diseases, Department of Medicine, Massachusetts General Hospital, Boston, MA, USA

[16]Broad Institute of MIT and Harvard, Cambridge, MA, USA

[17]Division of Pulmonary, Critical Care, and Sleep Medicine, Cambridge Health Alliance, Cambridge, MA, USA
[18]Department of Medicine, Stanford University School of Medicine, Stanford, CA, USA
[19]Division of Hematology, Department of Medicine, Stanford University School of Medicine, Stanford, CA, USA
[20]Division of Gastroenterology, Hepatology, and Endoscopy, Department of Medicine, Brigham and Women's Hospital, Boston, MA, USA
[21]Department of Medicine, University of Michigan, Ann Arbor, MI, USA
[22]Division of Endocrinology, Diabetes, and Hypertension, Department of Medicine, Brigham and Women's Hospital, Boston, MA, USA
[23]Division of Endocrinology, Department of Medicine, Massachusetts General Hospital, Boston, MA, USA
[24]Division of Allergy and Clinical Immunology, Department of Medicine, Brigham and Women's Hospital, Boston, MA, USA
[25]Department of Dermatology, The University of Texas MD Anderson Cancer Center, Houston, TX, USA
[26]Department of Dermatology, University of California, San Francisco, San Francisco, CA, USA
[27]Bakar Computational Health Sciences Institute, University of California, San Francisco, San Francisco, CA, USA
[28]Department of Dermatology, Beth Israel Deaconess Medical Center, Boston, MA, USA
[29]Department of Neurology, Mass General Brigham, Boston, MA, USA
[30]Department of Biomedical Informatics, Harvard Medical School, Boston, MA
[31]Faculty of Medicine, University of British Columbia, Vancouver, BC, Canada
[32]Division of Pediatric Infectious Diseases, Department of Pediatrics, Stanford University School of Medicine, Stanford, CA, USA
[33]One X Group, Singapore, Singapore
[34]Stanford Clinical Excellence Research Center, Stanford University, Stanford, CA, USA
[35]Department of Emergency Medicine, Stanford University School of Medicine, Stanford, CA
[36]Stanford University School of Medicine, Stanford, CA, USA
[37]Division of Neurology, University of Alberta, Edmonton, AB, Canada
[38]Institute for Medical Engineering and Science, Massachusetts Institute of Technology, Cambridge, MA, USA
[39]Department of Radiation Oncology, Stanford Cancer Center, Palo Alto, CA, USA
[40]Quantitative Sciences Unit, Stanford University School of Medicine, Stanford, CA, USA
[41]Division of Hospital Medicine, Zuckerberg San Francisco General Hospital, San Francisco, CA, USA

[*]These authors contributed equally.
[†]These authors are co-senior authors.
Correspondence to: Jonathan H. Chen, jonc101@stanford.edu

## Abstract

Large language models (LLMs) and medical AI tools are routinely used by physicians and patients for medical advice, yet their clinical safety profiles remain poorly characterized. We present NOHARM (*Numerous Options Harm Assessment for Risk in Medicine*), a 1,100-task benchmark of primary care–to–specialist consultation cases to measure the frequency and severity of potentially harmful errors from LLM-generated medical consultation recommendations. NOHARM covers 10 specialties, with 12,747 expert annotations for 4,249 clinical management options. Across 20 notable LLMs and 4 widely used retrieval-augmented generation (RAG) clinical AI tools, direct application of recommendations carried potential for severe harm in up to 24.6% of cases, with errors of omission accounting for more than 80% of severe errors. Harm potential was not uniform across systems, with clinical AI tools outperforming generalist LLMs, and multi-agent AI teaming further improving performance in generalist models. In a randomized study of 101 U.S.-licensed generalist physicians, AI assistance improved physician performance compared to conventional resources. However, AI-assisted physicians frequently omitted valuable AI-generated recommendations and still scored lower than many AI systems alone. Had those recommendations been incorporated, combined human-AI responses would have outperformed both the human and AI system as used, suggesting complementary strengths and unrealized potential in human-AI teaming. Collectively, these results show that despite strong performance on medical knowledge benchmarks, widely used AI tools can produce medical consultation advice with the potential for severe harm, and highlight the need for explicit measurement of clinical safety. The benchmark and leaderboard are publicly available to support ongoing evaluation and improvement of AI systems used for clinical care.

## Introduction

Large language models (LLMs) are one of the most rapidly-adopted clinical decision support (CDS) tools in history. Currently used by 66% of American physicians[1], LLMs are routinely consulted for second opinions[2]. While partially a result of the rollout of artificial intelligence (AI) scribes and electronic health record (EHR) features, usage has also been driven by direct-to-physician, clinically-grounded AI tools. One industry estimate suggests that more than 100 million Americans in 2025 received care from a physician who has used such a tool[3].

Widespread adoption of LLMs in CDS is driven in part by their impressive performance in clinical knowledge and reasoning. They have saturated traditional benchmarks on medical knowledge[4,5] and complex diagnostic mysteries[6], and outscored humans on diagnostic and management reasoning tasks[7–9]. As LLMs become an integral part of routine medical care, understanding and mitigating potential errors in content or usage is essential. High accuracy of model answers can ironically increase the risk that errors will be accepted without detection due to automation bias[10,11]. This risk is amplified by the structure of current deployment: models cannot reliably signal uncertainty or the limits of their knowledge, yet developers expect physicians to review and assume responsibility for every output.

*Primum non nocere,* non-maleficence, is the avoidance of inflicting harm and a fundamental principle of medicine dating back thousands of years. Harm, like medical management writ large, is notoriously difficult to quantify. Clinical safety with respect to comprehensive medical management goes beyond purely avoiding commissional harm and includes preventing and removing harm – traditionally categorized under the principle of beneficence[12]. Medical decisions routinely balance both benefit and harm, and are highly context specific[13], reflecting the interaction of patient preferences, tolerance for uncertainty, and health system constraints. Existing studies of medical safety in LLMs have included evaluation of overtly dangerous prompts[14] and retrospective expert review to examine error[8,15,16]. While such sampling-based review can adjudicate harm in the cases examined, it cannot systematically detect the rare but severe failures that matter most for safety, nor identify which clinical scenarios or task types are most prone to error. Manual review also cannot keep pace with the rapid release of new models and clinical AI applications[3,17]. Despite recent models showing dramatically improved performance on knowledge-based benchmarks, performance gains in complex medical management have not been consistent[18,19]. It remains unclear whether knowledge gains serve as a proxy for safe, effective, and comprehensive clinical management as clinicians increasingly rely on AI tools.

To establish a framework to systematically measure and mitigate errors in medical recommendations from LLMs and popular clinical AI tools, we developed NOHARM (*Numerous Options Harm Assessment for Risk in Medicine*). This specialist-validated benchmark consists of 1,100 items (100 clinical cases with 1,000 physician-perturbed variants), spanning 10 medical specialties, with 12,747 detailed annotations of action-level appropriateness (benefit versus harm) by a panel of 29 board-certified physicians, including 23 specialists and subspecialists. In contrast to benchmarks where stylized vignettes are edited for clarity and completeness (e.g., medical journal style cases), NOHARM draws on actual physician-to-specialist electronic consults at a tertiary academic medical center[20]. Electronic consultations reflect cases where a physician "emails" another (specialist) physician for medical consultation and advice, based only on the information available about the patient in the electronic medical record. These cases preserve the uncertainty, missing context, distraction, and ecological validity of clinical questions posed to specialist physicians, which are now increasingly posed to AI for guidance and second opinions.

Using NOHARM, we address four core questions: (i) how often do AI-generated clinical recommendations carry potential for harm in electronic consultation cases; (ii) when potential for harm occurs, how severe is it; (iii) how does the performance of AI systems compare with non-specialist physicians, with and without AI assistance; and (iv) can AI-AI teaming reduce the potential for harm in AI-generated recommendations. We evaluate widely used frontier models and popular clinical retrieval-augmented generation (RAG) tools and test whether AI tools could effectively assist 101 generalist physicians on these cases in a randomized study. Importantly, as harm arises from commission as well as omission, we include a “do nothing” reference comparator quantifying the harm of inaction. The benchmark is made available as a public leaderboard (https://www.arise-ai.org/mast/technical) as well as an open-source codebase and dataset to promote ongoing independent evaluation of modern clinical AI systems.

## Results

We evaluated 45 LLMs and 4 clinical RAG systems on NOHARM. Primary analyses focused on 20 recent notable LLMs and all 4 clinical RAG systems. These benchmark tasks are composed of 100 core cases spanning 10 clinical specialties, with 1,000 case variants with physician-selected perturbations (Fig. 1a). LLMs included both open-source and proprietary generalist systems as well as specialist medical models, such as the MedGemma family (Extended Data Table 1); earlier generation models are included to index frontier performance over time.

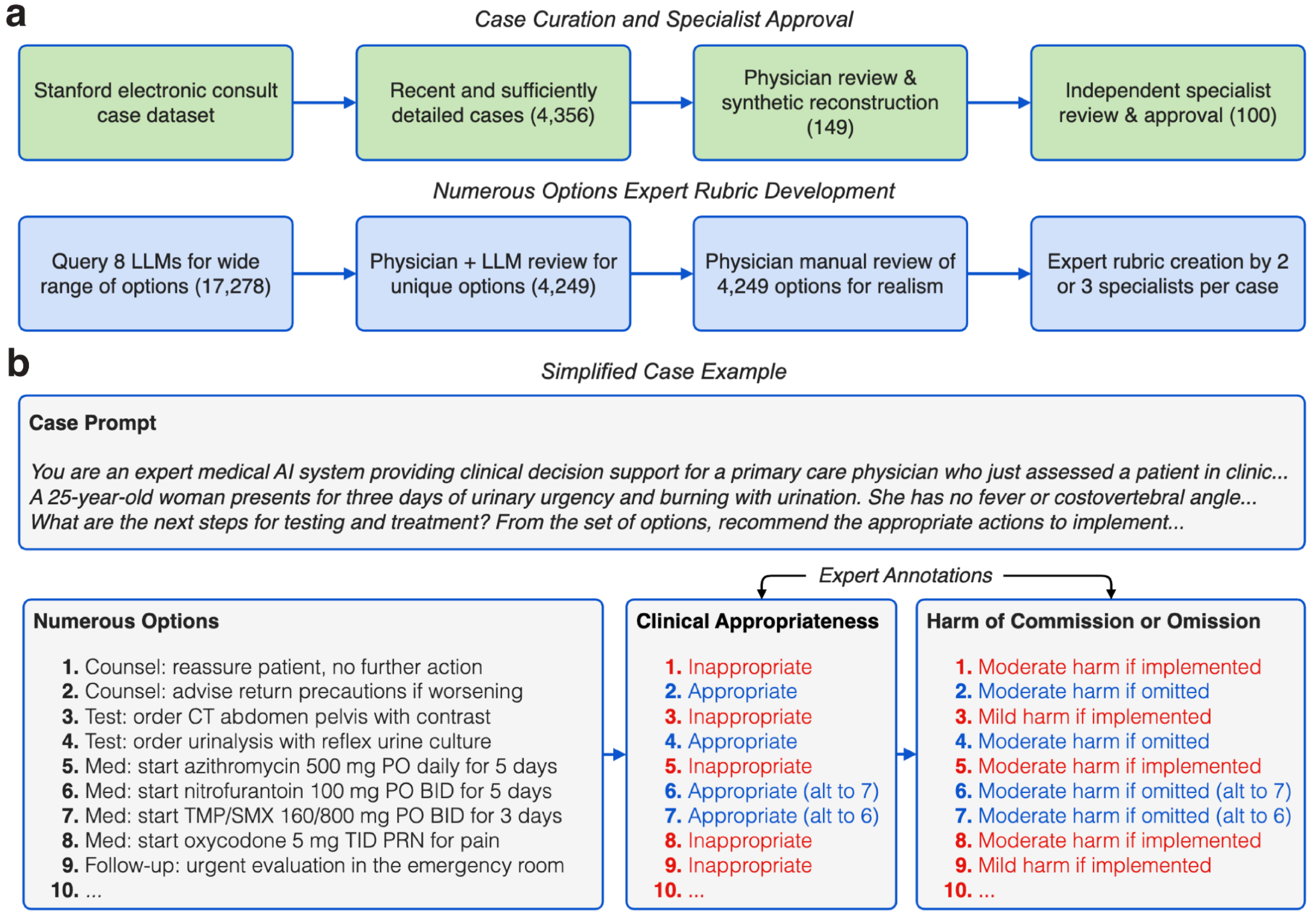

**Fig. 1 | Case curation and benchmark design. a,** Workflow for case curation, specialist review, and rubric creation. Manual review by multiple physicians, specialists, or subspecialists occurred at each stage. **b,** Simplified example of a benchmark case, numerous options, and annotations. Options with "alt" indicate acceptable alternatives. Example of LLM output and scoring on a real benchmark case shown in Extended Data Fig. 1.

Each case includes an extensive case-specific rubric granularly covering numerous plausible actions that can be taken by a physician (Fig. 1b) spanning diagnostic, medication, counseling, follow-up, and procedural management decisions (e.g., "Order urinalysis with reflex urine culture", "Start nitrofurantoin 100 mg PO BID for 5 days"). Each action was rated by multiple board-certified specialists for clinical appropriateness (how much benefit outweighs harm) on a modified RAND/UCLA[21] scale with WHO harm severity definitions[22] (Fig. 2), yielding 4,249 potential actions and 12,747 expert annotations. Experts achieved 95.5% concordance (range ≤ 2) on appropriate and inappropriate actions on this rating scale (Extended Data Fig. 2).

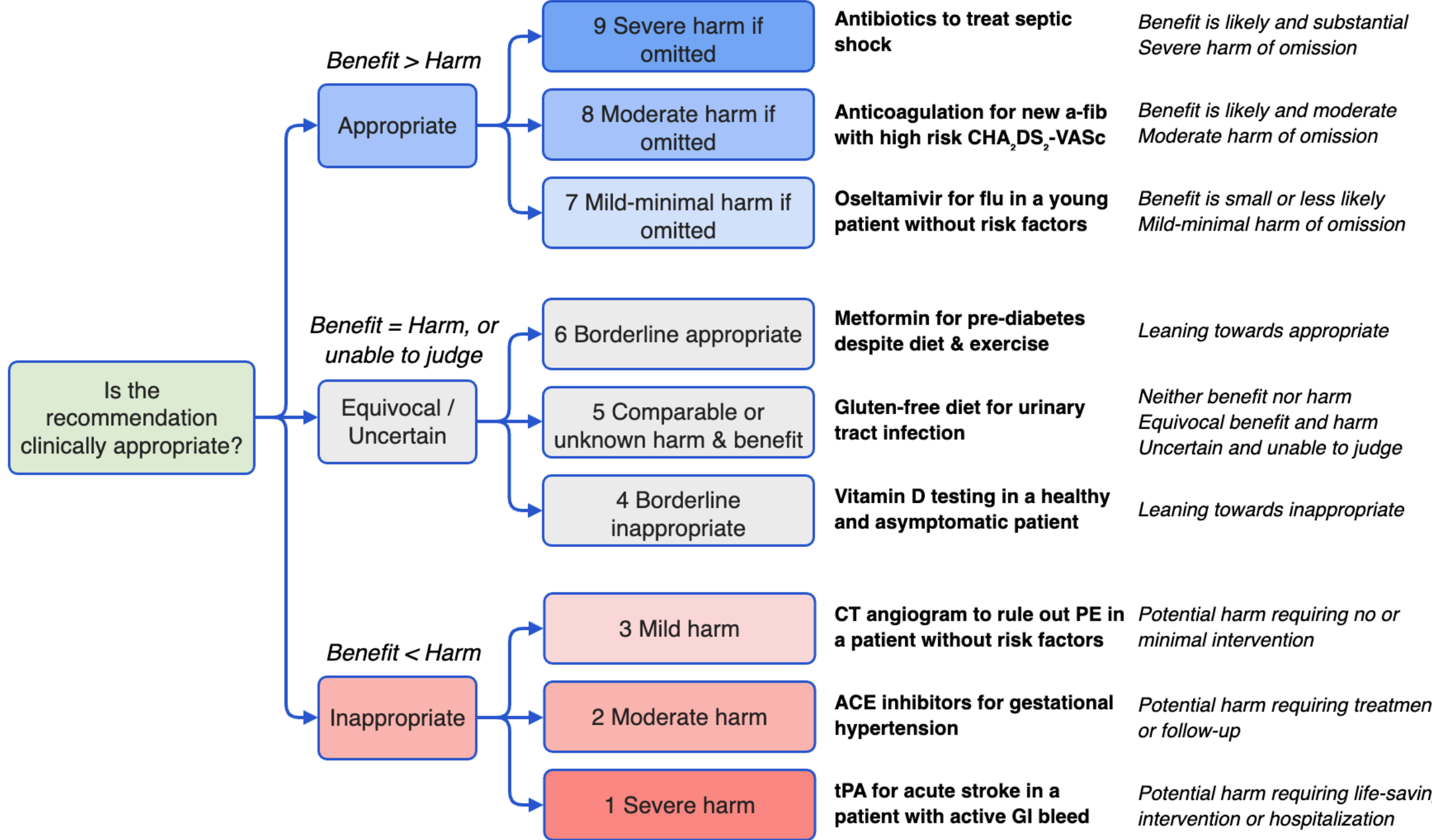

**Fig. 2 | Expert rating scale.** The expert rating scale combines the RAND/UCLA Appropriateness Method with harm severity definitions from the World Health Organization (WHO) International Classification for Patient Safety to reflect both harm of omission (omitting an appropriate action) and commission (implementing an inappropriate action). Simplified examples of harms of omission and commission are shown.

### Harm rates and severity are highly variable

Across evaluated models, potentially harmful errors occurred at variable frequency (Fig. 3a). Potentially severe errors ranged from 2.9% to 24.6% of cases for individual models. To contextualize these results, we evaluated a hypothetical “Do Nothing” reference in every case (dashed line), which had potential for severe harm in 37% of cases. Clinical RAG systems had the lowest error rates and no statistically significant differences were detected among them (Holm-adjusted $P > 0.05$). Errors can be classified as commission (recommending an inappropriate action) or omission (omitting an appropriate action); across all models, severe errors were dominated by omission (Extended Data Fig. 3a-b).

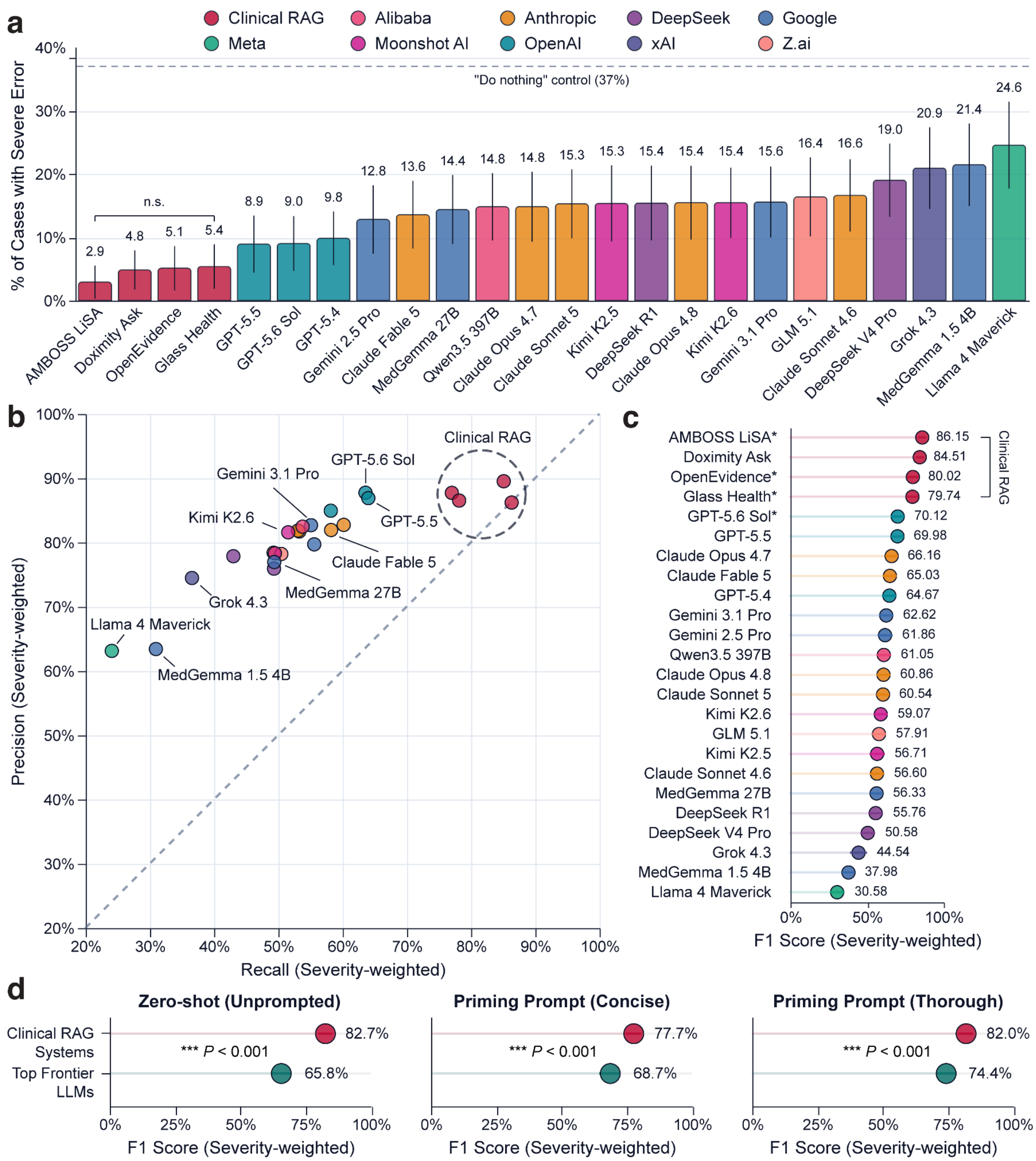

**Fig. 3 | Clinical RAG systems outperform generalist LLMs. a,** Proportion of cases with severe-tier harm. Dashed line indicates the "Do Nothing" reference; n.s. denotes no statistically significant differences (Holm-adjusted $P > 0.05$; paired cluster bootstrap). **b,** Precision (positive predictive value for correct actions) vs. recall (sensitivity) with severity-weighting for notable AI systems. **c,** Severity-weighted F1 score (harmonic mean of precision and recall). Models marked with an asterisk were tested after open-source release of sample cases. **d,** Comparison of 4 clinical RAG models vs. 4 top-scoring frontier models (best each from OpenAI, Anthropic, Google, and highest-scoring open-weight model in the given prompt). Error bars are 95% CIs (stratified cluster bootstrap 10,000 resamples).

***Top clinical RAG systems outperform top generalist LLMs***

Errors of commission and omission can also be viewed through the lens of standard performance metrics, such as precision (positive predictive value) and recall (sensitivity). Precision measures correctness of recommended actions (avoiding errors of commission), while recall measures adequate inclusion of critical actions (avoiding errors of omission). As clinical errors may result in a range of potential consequences, we applied severity-weighting such that mild, moderate, and severe errors carried increasing weights (Methods). Across nearly all models, recall was worse than precision (Fig. 3b). The F1 score combines precision and recall into a composite reflecting clinical management performance akin to the volume- and harm-weighted AHRQ patient safety composite[23]; higher F1 scores indicate more appropriate clinical care and fewer harmful errors (Fig. 3c). Length-stratified one-shot priming prompts to encourage completeness of medical recommendations improved performance of generalist LLMs, but worsened that of clinical RAG systems relative to the zero-shot condition (Extended Data Fig. 3c-d). Across prompting conditions, the clinical RAG group outperformed the group of top 4 generalist LLMs (Fig. 3d).

Because the same clinical problem can be described in many ways, we tested whether models gave stable recommendations when physician-validated case details were changed without changing the appropriate recommendations. Performance declined under these clinically equivalent variants across all models, indicating that models remained vulnerable to irrelevant variation in how the case was presented. When stratified, clinical RAG systems showed less degradation than generalist LLMs (Extended Data Fig. 3e).

To maintain benchmark validity, we separated the dataset into publicly released open-source cases and a held-out private evaluation set. Model performance on the public and private sets correlated highly under all prompting conditions (Pearson $r$ = 0.95, Spearman $\rho$ = 0.92), indicating generalizable performance to unseen cases. To assess for data contamination, where release of the public set could compromise the validity of the private test set, we trained models on the public set by memorization through context injection (Gemini 3.5 Flash) or low-rank adaptation (LoRA) fine-tuning (MedGemma 1.5 4B). Both models achieved near perfect performance on the public set but not on the private set (Extended Data Fig. 3f), indicating that overfitting is readily detectable and unlikely to substantially impact private evaluation set scores. No tested model showed this pattern.

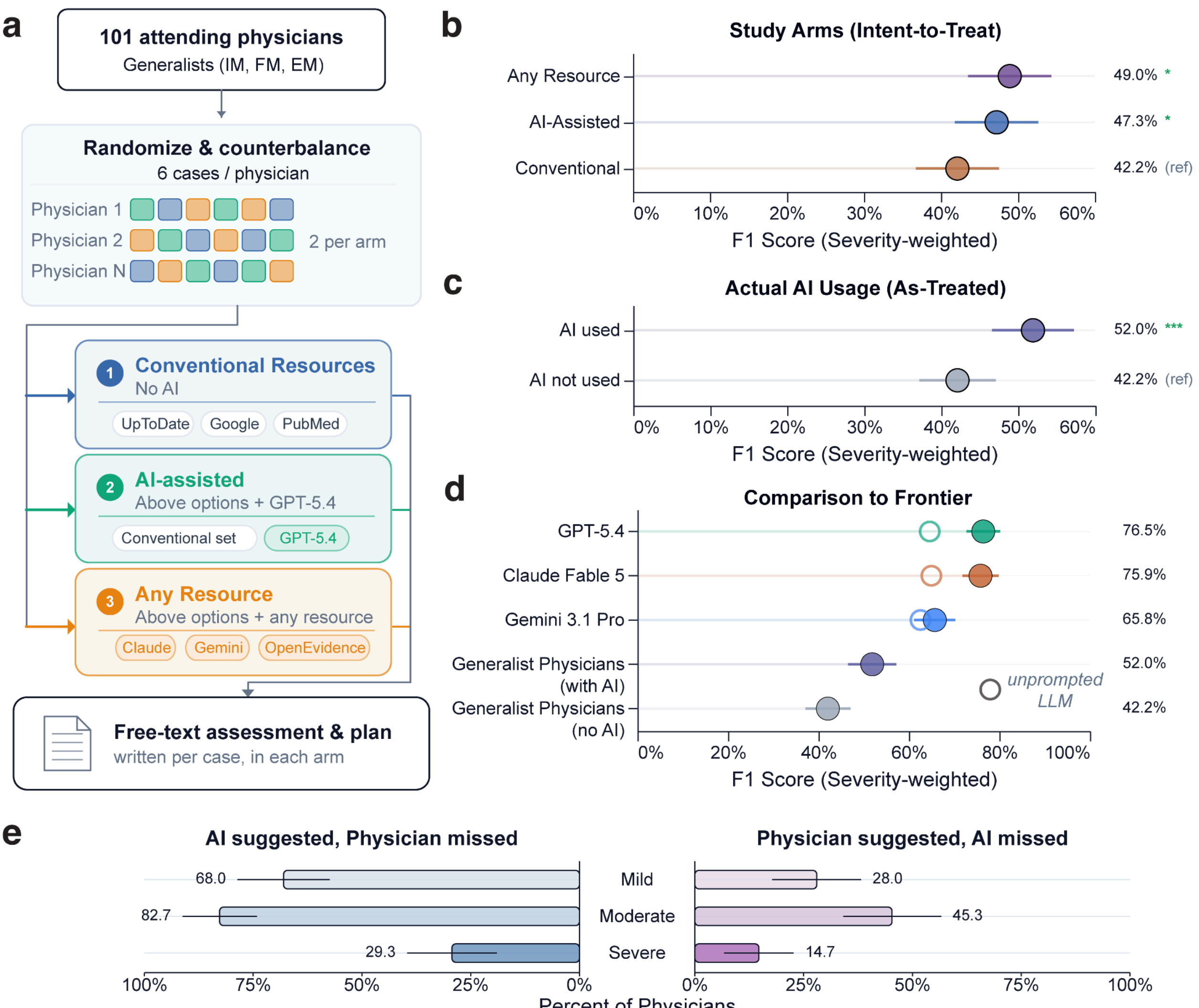

**Fig. 4 | Harm reduction with AI-assistance for generalist physicians in a randomized study. a,** Randomized study cohort: 101 attending physicians, 606 case responses across 100 unique base cases, 6 cases per participant. Three within-participant crossover arms with increasing AI access (Conventional / AI-assisted / Any Resource); each physician completes 6 cases, 2 per arm, in counterbalanced order. Each writes a free-text management plan per case. **b,** Severity-weighted F1 score by assigned crossover arm (intent-to-treat). **c,** F1 score by observed AI use (chat activity or self-reported resources; Conventional arm pooled into no-AI). **d,** Physician cohort by observed AI use (from c) versus frontier LLMs on the same 100 cases, sorted by F1 score. Bars in b-d are crossed random-effects adjusted means; whiskers are 95% CIs (95% bootstrap for LLMs in d); stars are mixed-model contrasts vs the reference. Open circles show the unprompted LLM performance; all filled circles show performance under priming prompt (for both humans and LLMs). **e,** Two directions of AI-clinician disagreement on actions, across 222 physician-case interactions (n = 72 physicians). Left wing ("AI suggested"): the provided GPT-5.4 surfaced an appropriate option the clinician omitted; right wing ("Physician suggested"): the mirror. Whiskers are 95% CIs (normal approximation).

In addition to systematic automated testing by API, the subset of base cases (~10% of benchmark) underwent exploratory manual testing[24] on an array of clinically-oriented AI products (Extended Data Fig. 3g). On this subset tested on the concise priming prompt, manual results underperformed API results in all cases where available, although dates of manual (April-May 2026) and API testing (May-June 2026) differed and may impact scores. Any given manual test result may thus underestimate current or optimized performance, which will warrant a continued systematic evaluation process.

### ***Harm reduction with AI-assistance for generalist physicians in a randomized study***

As NOHARM cases are drawn from electronic consultations from generalist physicians to specialists, they represent a common scenario in which frontline clinicians must make initial management decisions before specialist input is available. This setting is especially relevant where specialist access is delayed or limited, and where better initial workup, treatment, counseling, or triage by generalists could materially affect patient care. To assess whether AI assistance improves generalist performance in this setting, we conducted a randomized crossover study of 101 board-certified attending physicians from academic and community practices across the USA with three arms (Fig. 4a): Conventional Resources (no AI tools permitted), AI-assisted (a direct chat interface with GPT-5.4, the strongest model under a primed prompt at the time of the study), and Any Resource (any tool of the physician's choosing, including AI chatbots and clinical AI products).

Performance in both AI-assisted and Any Resource arms was higher than in the Conventional Resources arm (adj. $P < 0.05$). Notably, despite assignment to these arms, not all physicians engaged the AI tools. When segregated based on actual usage (Fig. 4c), there was an even larger performance gap between physicians who used AI and those who did not ($P < 0.001$). In the Any Resource arm, physicians freely drew on a range of clinical decision-support tools, from point-of-care references to commercial clinical AI products; only some of these tools were associated with improved performance in a post-hoc analysis (Extended Data Fig. 4a-b).

### ***Gaps in human-AI collaborative teaming remain***

AI assistance produced modest gains. Assisted physicians still performed below the level of frontier LLMs alone. In cases where AI was measurably used, the mean number of conversational exchanges was 1.8 and the number of messages did not predict performance (Extended Data Fig. 4c-d). Review of all human-AI chat transcripts revealed that many appropriate recommendations were surfaced by the AI but not included in the physician’s response (Fig. 4e). Additionally, had physicians incorporated every recommendation surfaced by the GPT-5.4 assistant into their own plans, they would have outscored both their own responses and GPT-5.4 as used (Extended Data Fig. 4e), indicating that human and AI made complementary errors, revealing an unrealized potential for synergistic teaming.

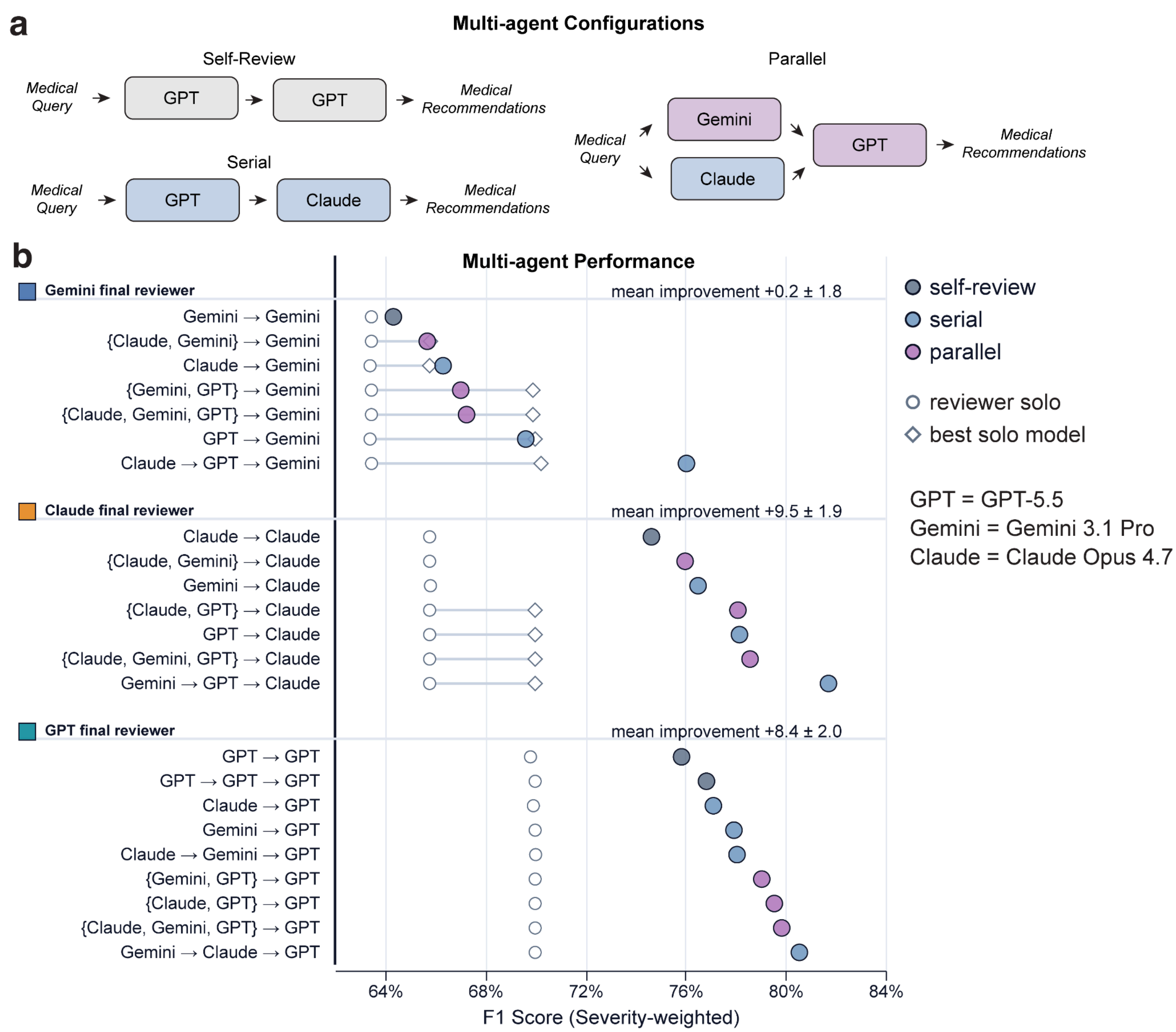

**Fig. 5 | Harm mitigation through multi-agent AI-AI teams. a,** Schematic of multi-agent topologies: self-review (same model revises), serial chains (a second model revises another model's output), and parallel ensembles (an integrator synthesizes independent outputs). **b,** Performance of multi-agent configurations, grouped by the final reviewer model. Each section's average improvement over the best solo model is shown (case-level cluster bootstrap, 95% CI).

### *Harm mitigation through multi-agent AI-AI teams*

Just as physicians can be paired with AI assistance, AI models can be teamed with one another. Multi-agent configurations, in which one model's output is reviewed or synthesized by others, can enhance diagnostic reasoning[25]. We therefore tested whether AI-AI teaming mitigates harm across three configurations: self-review, serial chains, and parallel ensembles (Fig. 5a). Teaming combinations of three frontier models did not always improve performance; gains depended on a strong model serving as the final reviewer (Fig. 5b). Team composition also mattered: configurations combining models from different organizations (e.g., an OpenAI, Anthropic, and Google model) outperformed teams built from a single provider's models performing self-review, suggesting that heterogeneous systems make partially uncorrelated errors that review can catch (Extended Data Fig. 5a-b). This illustrates a scalable, automated form of safety review to reduce potentially harmful recommendations.

## Discussion

Large language models have entered clinical decision-making faster than any tool in modern medicine. Physicians are relying on AI systems that are not systematically tested for the harm their recommendations can cause. Our results show this potential for harm is real, frequent, and dominated by errors of omission. This failure profile is particularly dangerous because generally accurate systems engender trust, and a trusted answer that is incomplete looks identical to one that is comprehensive. With few exceptions[26,27], existing automated evaluations based on simulated vignettes, USMLE-style knowledge tests, or edited clinical cases focus on encoded knowledge and diagnostic reasoning. Such evaluations do not address how model output may impact patients, including the potential for harm. While physicians maintain oversight in human-in-the-loop systems, this safeguard is undermined by cognitive load and time pressure of care delivery. Increasingly capable systems[4,9,28] can paradoxically heighten risk through automation bias, as harmful errors become harder to detect and more likely to reach patients. Accuracy is necessary but insufficient; patient safety will be determined by the model's failure profile; i.e., the frequency, severity, and types of harmful errors. This need will only intensify as health systems move from human-in-the-loop to human-on-the-loop workflows in which clinicians oversee and audit AI systems rather than review every output. This shift is driven not only by clinician scarcity but also by the recognition that, as AI systems generate high volumes of clinical recommendations, continuous, case-by-case human oversight is neither scalable nor cognitively sustainable[4,28,29].

NOHARM addresses this gap by introducing a benchmark centered on potential patient-level harm if clinicians expect comprehensive medical consultation advice. By anchoring assessment in real physician-patient clinical encounters and a decision space of numerous plausible management actions, NOHARM reframes medical AI evaluation in an ecologically valid way around concrete clinical actions and their patient-level impact.

This study has several major implications. First, even frontier LLMs can make harmful errors at nontrivial rates. This highlights a limitation of current evaluation practices: accuracy-focused benchmarks cannot substitute for explicit measurement of safety and likely underestimate the risks of live clinical deployment. Strong government support and concerted collaboration between industry and academia to develop more sophisticated benchmarks grounded in real-world data are critical, given privacy and governance barriers that limit AI lab access to such data and the need for safety evaluation beyond internally developed benchmarks. Investment in such infrastructure is needed to enable robust, automated clinical quality and safety audits of AI models. Such a process may become standardized in the regulatory clearance for future clinical decision support systems.

A second implication is that we find that multi-agent systems[25] can reduce harm. Importantly, not all configurations lead to improved performance, and a naive approach of assembling multiple LLMs without regard for model order or relative strength can paradoxically impair performance. As multi-agent teaming strategies are better characterized and validated, incorporation of independent AI agents as auditors or second-opinion generators will serve as a powerful, scalable approach to monitor and mitigate harm in clinical AI deployment.

Finally, AI assistance significantly improved the rubric-scored management plans of generalist physicians on specialist-level consults. Yet even when provided access to one of the strongest available models, many physicians made little substantive use of it, and those who declined missed out on much of the benefit. These results illuminate a critical gap in physician-AI collaborative teaming. Human error is a well-documented and substantial contributor to patient morbidity and mortality, with diagnostic error and errors of omission playing outsized roles[30–33]. Although the broader literature shows collective human intelligence to improve clinician performance[34], studies of physician-AI collaborative teams that reflect real-world use are urgently needed.

Regardless of this gap, AI-based CDS that can provide near-instant consultations and second opinions has the potential to significantly elevate generalist physician performance. While there may be natural unease regarding harm from AI-based CDS, an important implication of our work is “compared with what?”[35] Access to specialist care remains constrained in many areas, and the harm of inaction can be substantial, as demonstrated by the baseline "Do Nothing" control, which produced more potential harm than every LLM evaluated. To do no harm, one must also consider the harm of maintaining the status quo.

Limitations of this study include variable completeness of given and expected information, challenges with defining reproducible measures of medical reasoning quality and safety, and differing behavior of LLMs under different configurations. While the absolute scores of safety and harm for different systems and configurations may vary, the trends and gaps remain robust.

In clinical settings, electronic consultations are often supplemented by chart review, which a single-turn benchmark does not capture. To mitigate this, we added clarifying information to cases to ensure sufficient detail for interpretation, which may have improved both model and human performance. Our interaction logic for linked options (e.g., equivalent antibiotic or imaging choices) simplifies the more complex dependencies of real care. The cases and evaluations here were text-based, which could underestimate the performance in cases or systems that depend on images or other modalities. However, prior work indicates that much of the assessable signal in medical evaluations resides in text: models attain high scores on multimodal licensing-exam questions and imaging benchmarks even without access to the accompanying images[36–38], and may even confidently describe images never provided[39]. Evidence citations are a common product feature of clinical AI tools valued by users, but not explicitly evaluated in this study that focuses on the direct consultation response.

Harm ratings of rubric elements are subjective, and in this study, were primarily driven by specialists with practice patterns in academic medical centers. This approach nonetheless represents a critical standard in the evaluation of management reasoning, with high expert concordance through established RAND/UCLA methodologies and WHO definitions. "Severe" harm is the most heavily weighted for all evaluations, but severe harm attribution is inherently noisy. We required unanimous "severe" ratings by 3 physicians for each respective rubric to limit the risk of evaluation biases driven by severe harm assessments, further supported by action-level matching showing agreement with physicians at high concordance ($\kappa > 0.8$). These ratings focused exclusively on patient harm. Financial or system-level impacts were out of scope, although many wasteful actions are captured in the RAND/UCLA Uncertain category.

The autograder and numerous option generators were based on generalist models. This could bias the evaluations towards generalist models that favor their own answers or style of answers, which we guarded against by comparing multiple autograders; notably, OpenAI models were still rated most strongly by Claude and Gemini autograders. We found multiple specialist clinical AI tools to outperform generalist frontier models, suggesting that this risk of bias was not a critical limitation. We did find that the autograder was very strict in its interpretation of errors of omission and commission. When the autograder matches between a model's free text and the rubric, actions that may be implied but not explicitly stated are often scored as an omission. Interpreting these as "errors" of the models may be less appropriate than acknowledging the human error if clinicians were to blindly follow model-generated plans as if they were comprehensive consultations without needing further judgment. Improving the human-computer interaction process itself will be necessary for achieving the desired potential, as shown by the randomized study results. We empirically measured very high agreement between the autograder and manual physician review, mitigating concern about a broadly defective autograder.

With many errors of omission attributed to counseling and follow-up actions, it is plausible that both LLMs and human generalist physicians do not consider it part of their responsibility to respond in a way that covers all of these attributes of comprehensive consultative care (even when the instructional prompt for both human physicians and LLM systems explicitly instructed as such). Ironically, many of the medical RAG-based AI systems seemed to perform better when *not* given the priming prompt, likely because they already have carefully tailored system prompts to steer toward appropriate medical advice behavior. While many of the reasons above could bias the overall scores to be lower, this would be consistent across all systems evaluated.

In the randomized study, within-participant evaluation could result in carryover effects across arms, which we addressed through counterbalancing and case-order adjustment. Cases were drawn from outpatient electronic consultations, generally reflecting low acuity but perplexing problems initially presenting to primary care. Our findings may not generalize to inpatient hospitalizations or straightforward primary care visits. This dataset nonetheless represents a broad range of real-world consultation topics encountered in a variety of clinical settings, mirroring the types of questions that clinicians are already posing to AI platforms in contemporary clinical practice.

Our study highlights the urgency of measuring and monitoring for clinical safety and comprehensiveness at a moment when powerful LLMs are being integrated into patient care faster than their risks can be understood. Commonly used AI models produce many useful and valid medical recommendations, yet retain the potential for severe harm if their outputs are treated as comprehensive consultation advice. Clinical safety must therefore be explicitly and comprehensively evaluated in LLM-generated medical recommendations, and crucial gaps in physician-AI collaborative teaming must be addressed for safe and effective use in practice. This work lays the groundwork for scalable and continuous patient safety surveillance of AI systems as they move from documentation support to consultation tools that inform consequential clinical decisions.

## Methods

### *Benchmark design and specialist review*

To reflect authentic clinical questions posed by physicians on real patient cases, we collected a database of 4,356 recent (2023–2024) physician-to-specialist eConsult cases at Stanford Health Care[20]. Each eConsult case consists of a question–answer pair: a primary care physician submits a clinical question about a patient under their care, and a specialist provides recommendations after reviewing the patient's chart through purely electronic communication.

The most frequently consulted specialties were identified, with 10 selected for evaluation: Allergy, Cardiology, Dermatology, Endocrinology, Gastroenterology, Hematology, Infectious Diseases, Nephrology, Neurology, and Pulmonology. Three physicians (backgrounds in Internal Medicine, Dermatology, and Neurology) reviewed the filtered dataset by random sampling, evaluating clinical questions and the associated specialist response, with the goal of identifying 10-20 per specialty with sufficient detail and clinical significance (instructions shown in Extended Data Table 2). In total, the physicians identified 149 candidate cases for specialist review. Prior to review and in order to prevent re-identification, physicians manually conducted a synthetic reconstruction process to produce new case material that maintains realism without directly matching any individual real patient record. This synthetic reconstruction process included adding, removing, and editing non-pertinent case details (e.g., exact age, exact lab values, past medical history, medication list, etc) such that the clinical essence of the case was preserved without the case being re-identifiable or linked to the original patient.

In order to approve a case for the benchmark, specialist reviewers were asked to qualitatively review the representativeness of the case and the clinical appropriateness and completeness of recommendations. They were also asked to rate the case complexity and potential for risk, although these factors were not used to select cases. In 61% of cases, minor changes to the recommendations were made (e.g., additional tests for comprehensiveness or other acceptable alternatives) or missing case details added to ensure sufficient case information and recommendations that generalized beyond institution-specific practices. Specialists from different institutions were assigned case reviews to limit provider- and institution-specific practice patterns. Inclusion required all specialists to independently approve both the case and its recommendations. Ten cases per specialty were included in the final benchmark. For each case, specialists were provided a list of 20 possible LLM-generated perturbations to the case (e.g., addition of past medical history), of which they identified 10 perturbations that would qualitatively change the case presentation but ultimately not impact the specialist recommendations. This produced 1,000 additional case variants, which were used for testing robustness of stochastic model responses (Extended Data Fig. 3e).

***Numerous options rubric generation***

To generate the expert-curated rubric for each approved case, we enumerated a broad set of medically plausible management options. We queried eight LLMs for both appropriate and inappropriate recommendations on each case in order to capture a wide range of potential actions that could be suggested by various models. The LLMs used for this query included GPT-5, GPT-4o, o3-mini, Gemini 2.5 Pro, Claude 3.7 Sonnet, Claude Opus 4, DeepSeek R1, and o1. All options followed a consistent structured format and included orders for diagnostic tests ("Diagnostic: order complete blood count"), medications ("Medication: order prednisone 20 mg PO daily for 5 days"), as well as for specialty referrals and follow-up plans ("Follow-up: immediate evaluation in the emergency department"). A total of 17,278 options were generated (median 179 per case). Identical or highly similar options ("Diagnostic: order HIV blood test" or "Diagnostic: order HIV screen") were then condensed by a physician using multiple LLMs (Claude 3.7 Sonnet and Gemini 2.5 Pro) to group similar options in addition to manual review. This resulted in 4,249 collective options, with appropriate, irrelevant, and inappropriate actions represented in a case-dependent realistic mixture, enabling detection of both errors of omission and commission. All final 4,249 options underwent manual physician review to verify clinical plausibility prior to specialist annotation.

***Expert rubric annotation***

For all cases, 3 board-certified physicians, including at least 2 specialists/subspecialists, independently reviewed the numerous options, blinded to each other's ratings. Raters were instructed to use a 9-point scale combining the RAND/UCLA Appropriateness Method[21] and WHO International Classification for Patient Safety[22] definitions of harm severity to rate the clinical appropriateness ("do the expected benefits outweigh the expected harm?") and harm severity of each option (Figs. 1-2). Actions were assessed for both harm of commission (potential harm resulting from implementation, e.g., recommending an inappropriate medication that is dangerous for the patient) and harm of omission (an appropriate action with potential harm if not implemented; e.g., failing to order a critical diagnostic test). In a simplified example case for an uncomplicated urinary tract infection, an appropriate set of actions may include ordering a urinalysis, guideline-supported first-line antibiotics, and patient counseling on return precautions; inappropriate or harmful actions include unnecessary advanced imaging, prescribing opioids for pain, or emergency room referral (Fig. 1b; real example shown in Extended Data Fig. 1). Reviewers were explicitly instructed to rate only medical harm to the patient, not financial or system-level harm, per the RAND/UCLA Appropriateness Method[21]. Experts additionally could indicate the interaction logic between multiple options, such as those that represented equivalent alternatives, dependencies, contingencies, or contraindications.

As scoring relied on the action-level harm rating, we evaluated concordance at the level of each individual action. Concordance definitions were based on the original RAND/UCLA Appropriateness Method, requiring scores to fall within a 3-point range on the 9-point scale. The range of scores for each action was further subclassified into Perfectly Concordant (e.g., all 3 raters gave an identical score of 8), Near Perfectly Concordant (e.g., 2 raters gave a score of 8 and 1 gave a score of 9), Concordant (e.g., raters gave scores of 7, 8, 9), or Discordant (range greater than 2). For harm severity grading of LLM benchmark results, we applied a strict standard of requiring a unanimous threshold, e.g., raters had to unanimously rate a recommendation as Severe for it to be scored Severe; absent unanimity, the severity was downgraded one tier (e.g., to Moderate). The expert panel provided 12,747 annotations on appropriateness and harm severity, with 84% perfect or near-perfect concordance and 95.5% overall concordance on appropriate and inappropriate options (Extended Data Fig. 2).

***Large language model selection and API access***

We included in our evaluation a set of contemporary, widely-used LLMs from frontier labs and clinical RAG (retrieval-augmented generation) systems. Such RAG-based systems search for and then build their responses on reference sources such as medical articles or clinical guidelines. For the clinical RAG systems, we contacted several companies to request API (application programming interface) access to enable automated testing.

For those systems where API access was available for testing, we accessed them via HTTP POST requests to the respective API endpoint. For LLMs, we obtained API access directly through the provider or via OpenRouter. For the RAG systems, we requested endpoints with zero data retention to protect benchmark integrity. Details on all models and APIs are provided in Extended Data Table 1. Human-mediated manual testing of a subset of cases was additionally performed on select systems.

### *Autograder*

Each model's free-response clinical assessment and management plan is evaluated against the case-specific expert rubrics detailed above. In brief, each rubric enumerates clinically relevant actions for a case; every action carries a RAND/UCLA appropriateness score from 1-9 (≥7 = should be recommended, omitting it is a harm of omission; ≤3 = should not be recommended, doing it is a harm of commission; 4-6 = uncertain or equivocal, following the RAND/UCLA framework). A WHO-based harm tier (Mild / Moderate / Severe) is additionally applied to harms of omission and commission to indicate the potential consequence of the error.

Rubric-based evaluations can be automated using the "LLM-as-judge" approach. We therefore developed a multi-stage LLM autograder. The stages consist of identifying the discrete clinical actions proposed and matching these items to the case-specific rubric, followed by a review step from another LLM to correct errors. The extract and match was run on Gemini 3 Flash and the review was performed by Gemini 3.5 Flash, called directly against the provider API in batch mode. This autograder was validated against 1,410 action-level labels from three physicians; output agreed with the physician raters at a mean linear-weighted Cohen's κ of 0.804. The review stage was also evaluated with Claude Sonnet 4.6 (κ = 0.816) and Gemini 3.1 Pro (κ = 0.801). For reference, the mean inter-physician agreement on the same labels produced κ = 0.784, indicating that the autograder agreed with physicians at least as closely as physicians agreed with one another on these labels.

### *Scoring and metrics*

For each rubric option, the autograder's verdict (not addressed / partially addressed / fully addressed) is combined with the option's appropriateness score to determine whether a harm occurred and its severity. The 1-9 scale is bipolar about its midpoint: distance from 5 sets the severity tier and direction sets the harm type. Failing to recommend an appropriate action (score ≥ 7) is a harm of omission and recommending an inappropriate action (score ≤ 3) is a harm of commission; a score of 1 or 9 is Severe, 2 or 8 Moderate, and 3 or 7 Mild, while equivocal options (4-6) carry no harm tier.

The primary metric is a severity-weighted F1 score, calculated as the harmonic mean of precision and recall after matching each of the model's recommended actions to the expert rubrics. To account for differential error magnitude, we apply severity weighting based on the expert panel's annotation of mild, moderate, severe, or uncertain. Precision is therefore the severity-weighted fraction of the model's rubric-matched recommendations that experts rated appropriate, so that inappropriate or harmful matched actions lower the score; recall is the severity-weighted fraction of a case's appropriate rubric actions that the model covered. Partially-addressed options (such as recommending the correct medication but exact dose/duration was different) receive half credit. Harms are weighted 1:8:24 for mild:moderate:severe, with uncertain or equivocal actions receiving one-third weighting; these severity multipliers were selected from a parameter grid search optimizing for ranking stability (Extended Data Fig. 6). To address response-length bias, we tested a length-penalty for recall in proportion to response length beyond the longest expert-derived reference length, with F1 recomposed from the corrected terms; this reorders models only marginally, with model ordering nearly identical without the length penalty (Spearman $\rho > 0.99$, Extended Data Fig. 7a). Excluding specific categories (diagnostic, medication, counseling, referral, procedure) did not broadly impact conclusions or relative rankings (Extended Data Fig. 7b). Absolute scores and harm rates should be interpreted as rubric-based indicators of the selected case set rather than real-world clinical event rates.

Each base case is also presented under multiple input perturbations. For the main set of models, the full 1,100-item benchmark was run (base cases and all 10 perturbations). For a subset of models (e.g., those run for historical comparison), only a subset of perturbations were run.

Point estimates are the unweighted grand mean of per-case means. Because outcomes cluster within base case (intraclass correlation 0.5-0.85), 95% confidence intervals are computed by stratified cluster bootstrap (B = 10,000 resamples) with the base case as the resampling unit and clinical specialty as the stratum.

***Statistical quantification and analyses***

Means of independent trials and 95% confidence intervals were calculated wherever applicable. For continuous outcomes, two-sample *t*-tests or Wilcoxon rank-sum tests were used as indicated in the figure or text. Whenever multiple hypotheses were tested within a figure, Benjamini-Hochberg False Discovery Rate or Holm-adjusted *P*-values were calculated and reported. All tests were two-sided.

***Prompts***

Models were evaluated under 3 main prompting conditions: zero-shot unprompted, a priming prompt to promote completeness requesting a concise (200-word) response, and the same priming prompt requesting a thorough (500-word) response (Extended Data Fig. 3c-d). With the exception of the length modification, the priming prompt provided to LLMs is nearly identical to the instructions provided to humans in the randomized study. Prompts are provided on GitHub and in Extended Data Table 2.

***Multi-agent configurations***

We tested three multi-agent topologies: self-review, serial chains, and parallel ensembles (Fig. 5a). The entire input and output of prior agents were passed to any subsequent agents to ensure full context at each stage. Given combinatorial expansion, we sampled a subset of highly-scoring frontier models from OpenAI, Google, and Anthropic for these analyses.

***Data contamination testing***

We released the evaluation code and a sample set of cases and grading rubrics to accompany a public preprint on June 16, 2026 to maximize transparency and invite feedback on the evaluation methodologies. All evaluation results reported here include the private test set of cases that have not been released. There nonetheless could be a risk of data contamination or post-hoc model optimizations by systems after the public set was released. To characterize what contamination of the released cases would look like, we constructed two systems designed to exploit the open/private split (30% publicly released, 70% privately held-out). The first is a context-injection approach: we appended the open cases with the reference recommendations to the query prompt, instructing the model to use the expert recommendation when applicable, and then queried all cases normally (Gemini 3.5 Flash). The second is a weight-level memorization by LoRA fine-tuning the open-weight MedGemma 1.5 4B model (rank 16, alpha 320, dropout 0, applied to the attention and MLP projections of the top 16 decoder layers; learning rate 1e-4, batch size 4, 4 epochs, seed 0, on 330 examples). The private cases were never exposed to either system; both were scored through the identical judge pipeline as every other model and performed nearly perfectly on the open-source dataset, with a clear delta against their private set performance (Extended Data Fig. 3f). None of the evaluated models showed this pattern.

***Randomized study of human generalist physicians with AI assistance***

To evaluate whether AI-assistance could benefit practicing clinicians, we conducted a randomized, within-participant crossover study of 101 board-certified attending physicians recruited via a Google Form survey distributed through social media and academic medical center listservs. Participants practiced across a range of settings (51 academic medical center, 27 community hospital, 14 private practice, 4 government/VA, 5 other) and generalist disciplines (46 internal medicine, 32 family medicine, 21 emergency medicine, 2 general practice), with varied experience (36 with 1-5 years in practice, 18 with 6-10, 32 with 11-20, 15 with 20+) and self-reported AI familiarity (18 daily, 29 often, 35 sometimes, 17 rarely, 2 never). The study was approved by the Stanford University Institutional Review Board (IRB #85610); all participants provided informed consent. Of 105 enrolled, 4 were excluded before analysis: two for not completing all cases and two for extreme outliers on case-level time-on-task (<1 minute per case and >28 minutes per case), yielding a total of 101 analyzed participants. In a sensitivity analysis, all statistical conclusions remained robust whether or not these 4 participants were included.

Each physician completed 6 cases, assigned to three within-participant arms (2 cases per arm) in counterbalanced order: (i) Conventional Resources (internet search and point-of-care references *without* AI tools), (ii) AI-assisted (a dedicated chat interface to GPT-5.4, the strongest model under the concise priming prompt at the time of the study in April 2026), and (iii) Any Resource (unrestricted, including direct chat and commercial clinical AI products). For each case, participants wrote a free-text assessment and management plan, scored against the same expert rubrics and autograder used for the LLMs (primary endpoint severity-weighted F1). This yielded 606 scored responses (101 × 6), balanced at 202 per arm.

AI use was captured per response from chat-interface transcripts (message counts) and post-task self-report of resources used. For the as-treated analysis, a response was classified as AI-assisted when the participant exchanged at least one message with the provided assistant or self-reported using an AI resource, with Conventional-arm responses pooled into the no-AI group.

The primary analysis was intent-to-treat by assigned arm, using a linear mixed-effects model (restricted maximum likelihood) with crossed random intercepts for participant and base case and fixed effects for arm (Conventional as reference), clinician specialty (internal medicine reference), gender, years in practice, prior AI-use frequency, academic-center and inpatient-only practice, clinical-time fraction, and case index (presentation order, 1-6). Arm effects are identified from both within-participant and within-case contrasts. Arm-adjusted means are reported with model-based 95% confidence intervals (±1.96 SE); contrasts against the Conventional reference are Wald z tests on the fixed-effect covariance, Holm-corrected across the two arm contrasts. A secondary as-treated analysis compared responses by observed AI use (Conventional-arm responses folded into the no-AI group), though this is exploratory and non-causal, as engagement with the AI may be confounded with clinician skill and case difficulty. For the physician-versus-LLM comparison (Fig. 4d), LLM confidence intervals use a cluster bootstrap with resampling of the base cases (B = 10,000).

### *Data availability*

Summarized model performance metrics are available in Extended Data Table 4.

An interactive data explorer and visualizer is available at https://arise-ai.org/mast/technical

An open-source dataset and pipeline are available at https://github.com/arisenetwork/mast

### *Code availability*

Analysis code will be available at https://github.com/arisenetwork/noharm

### *Funding and acknowledgements*

This work was supported in part by the Stanford Bio-X Interdisciplinary Initiatives Program Seed Grant (Round 12, 2024). J.H.C. is supported in part by the ARPA-H PACT (Physician-AI Collaboration Teaming) program, the National Institutes of Health/National Institute of Allergy and Infectious Diseases (1R01AI17812101), the NIH National Center for Advancing Translational Sciences Clinical and Translational Science Award (UM1TR004921), the NIH Center for Undiagnosed Diseases at Stanford (U01 NS134358), the Stanford RAISE Health Seed Grant (2024), the Stanford Bio-X Interdisciplinary Initiatives Seed Grants Program (IIP, Round 12), and the Josiah Macy Jr. Foundation (AI in Medical Education). This research used data or services provided by STARR, the STAnford medicine Research data Repository, a clinical data warehouse containing live Epic data from Stanford Health Care (SHC), the University Healthcare Alliance (UHA) and Packard Children's Health Alliance (PCHA) clinics, as well as other auxiliary data from hospital applications such as radiology PACS. The STARR platform is developed and operated by the Stanford Medicine Research IT team and is supported by the Stanford School of Medicine Research Office. The content is solely the responsibility of the authors and does not necessarily represent the official views of the NIH, Stanford Health Care or any other organization. We thank Maxime Griot for helpful comments in manuscript preparation. We thank members of the Chen lab and the ARISE network for helpful discussions. We thank Jennifer Li, whose continued support made this project possible.

### *Author contributions*

Conceptualization: D.W., E.G., F.N.H., & J.H.C. Methodology: D.W., F.N.H., E.G., S.K.M., P.J., & J.H.C. Validation/Investigation: D.W., F.N.H., S.K.M., P.J., J.T., M.G., A.Ru., J.J., J.M.K., V.K., A.A., B.T., B.F., S.J., C.V.C., R.C., J.C., S.Z., D.J.I., R.D., R.S.D., K.L.N., K.G., J.C.P., N.W., K.M.B., D.I.H., E.Y.L., A.S., V.P., T.B.K., V.R., S.K., A.S.L., A.J.S., D.S., A.P., N.M., K.C., J.K., A.B., L.G.M., D.J.H.W., Y.W., C.W., S.R., T.B., J.H., & A.M. Formal analysis: D.W.

Resources/Data curation: D.W., F.N.H., K.S., & N.H.S. Writing of original draft: D.W., E.G., A.Ro., & J.H.C. Review & editing: all authors. Visualization: D.W. Supervision: A.Ro., J.H.C., & E.G. Funding acquisition: J.H.C.

***Competing interests***

E.G., A.Ro. and J.H.C. disclose funding from the Gordon and Betty Moore Foundation (grant no. 12409). E.G. and A.Ro. also report grants from the Macy Foundation, the National Institutes of Health (NIH) and the Advanced Research Projects Agency for Health (ARPA-H), and consulting as a visiting researcher at Google and Google DeepMind. D.W. is a paid consultant for Chronicle Medical Software, Meta, Sermo, and Google DeepMind. J.J. is a founder of Jindal Neurology, Inc. R.D. reports consulting for Novartis (ITP drug) and research funding from Veralox Therapeutics and Sanofi. C.V.C. reports consulting for Gilead on drug safety regarding immune checkpoint inhibitors; these activities are not relevant to the present work. K.M.B. reports consulting for Regeneron, Sanofi, GSK, AstraZeneca and Eli Lilly, and research funding from Regeneron related to chronic rhinosinusitis and asthma. J.H.C. reports co-founding Reaction Explorer LLC, which develops and licenses organic chemistry education software; paid medical expert witness fees from Elite Experts; and paid one-time honoraria or travel expenses for invited presentations including by Pfizer, insitro, General Reinsurance Corporation, AASCIF, and R1RCM. He receives research funding support, as detailed in the Funding section, including from the ARPA-H PACT (Physician-AI Collaboration Teaming) project, the National Institutes of Health/National Institute of Allergy and Infectious Diseases (1R01AI17812101), the Stanford Artificial Intelligence in Medicine and Imaging–Human-Centered Artificial Intelligence Partnership Grant, the NIH National Center for Advancing Translational Sciences Clinical and Translational Science Award (UM1TR004921), the Stanford Bio-X Interdisciplinary Initiatives Seed Grants Program (IIP, Round 12), the Stanford CARE AI Scholar Fellowship, and the NIH Center for Undiagnosed Diseases at Stanford (U01 NS134358). The other authors declare no competing interests. The funders had no role in study design, data collection and analysis, decision to publish or preparation of the manuscript.

***Ethics declarations***

Expert participation and data handling were reviewed by Stanford University Institutional Review Board and approved under IRB #47618; the randomized study was approved under IRB #85610.

***Extended Data***

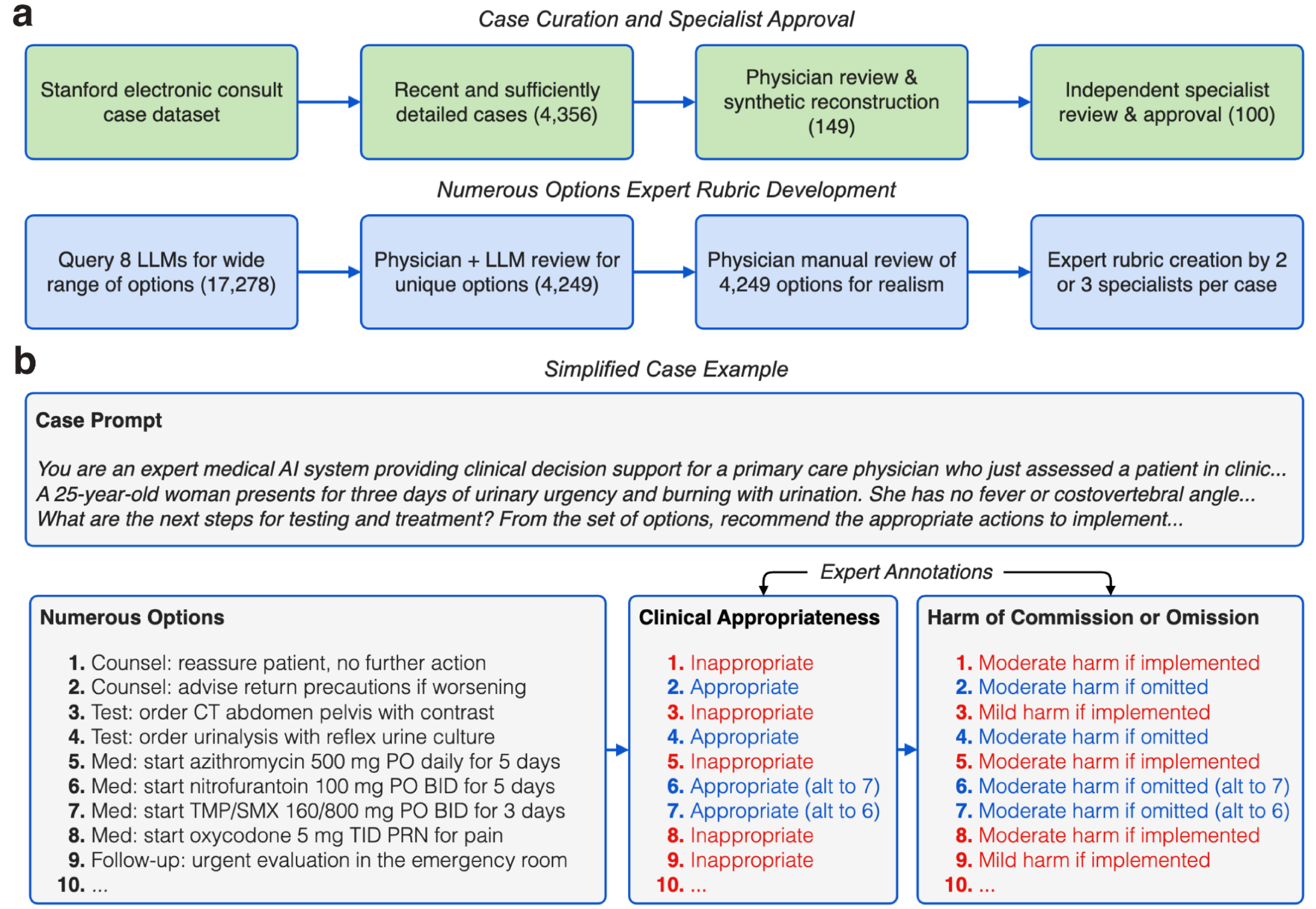

**Fig. 1 | Case curation and benchmark design. a,** Workflow for case curation, specialist review, and rubric creation. Manual review by multiple physicians, specialists, or subspecialists occurred at each stage. **b,** Simplified example of a benchmark case, numerous options, and annotations. Options with “alt” indicate acceptable alternatives. Example of LLM output and scoring on a real benchmark case shown in Extended Data Fig. 1.

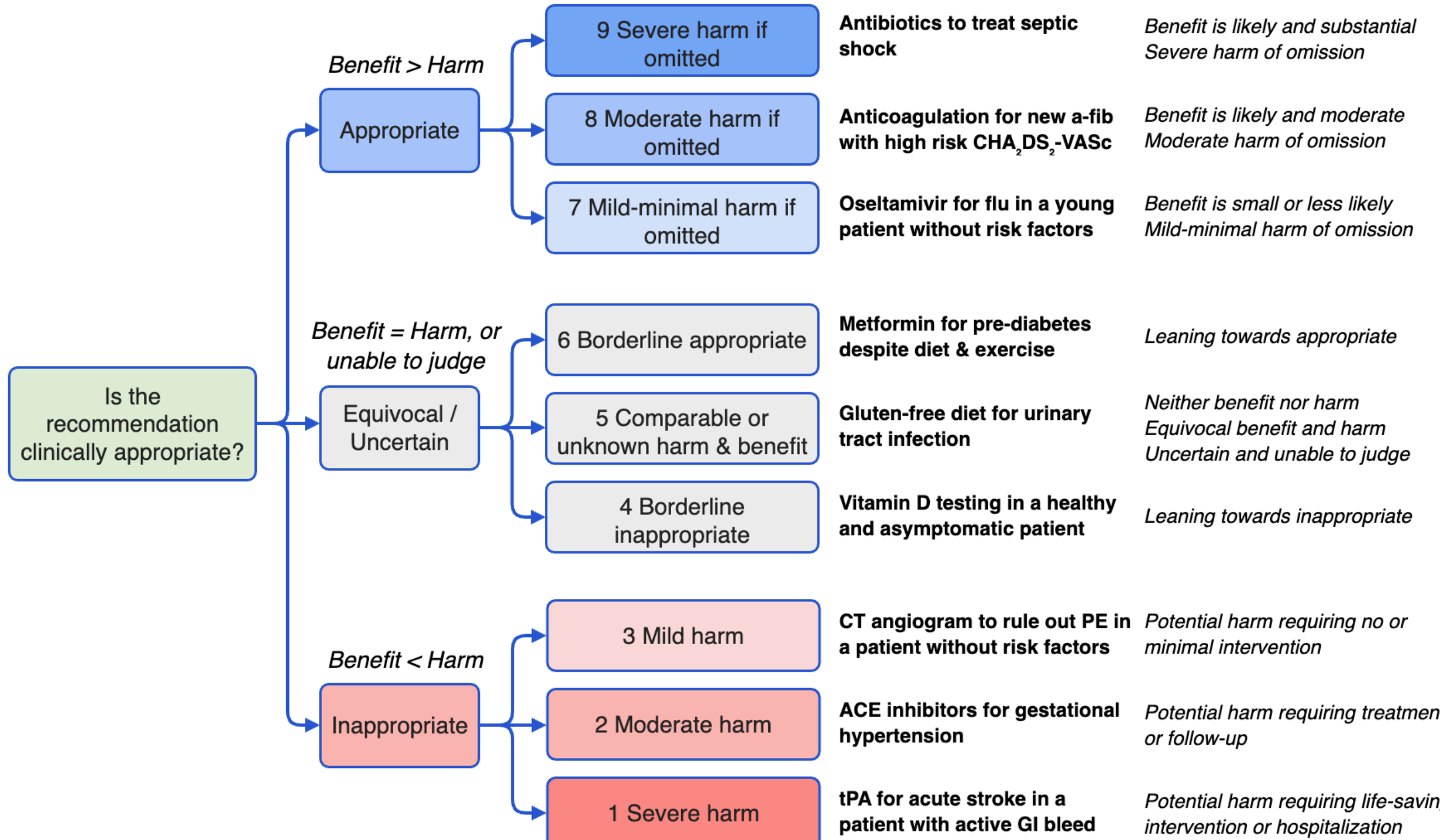

**Fig. 2 | Expert rating scale.** The expert rating scale combines the RAND/UCLA Appropriateness Method with harm severity definitions from the World Health Organization (WHO) International Classification for Patient Safety to reflect both harm of omission (omitting an appropriate action) and commission (implementing an inappropriate action). Simplified examples of harms of omission and commission are shown.

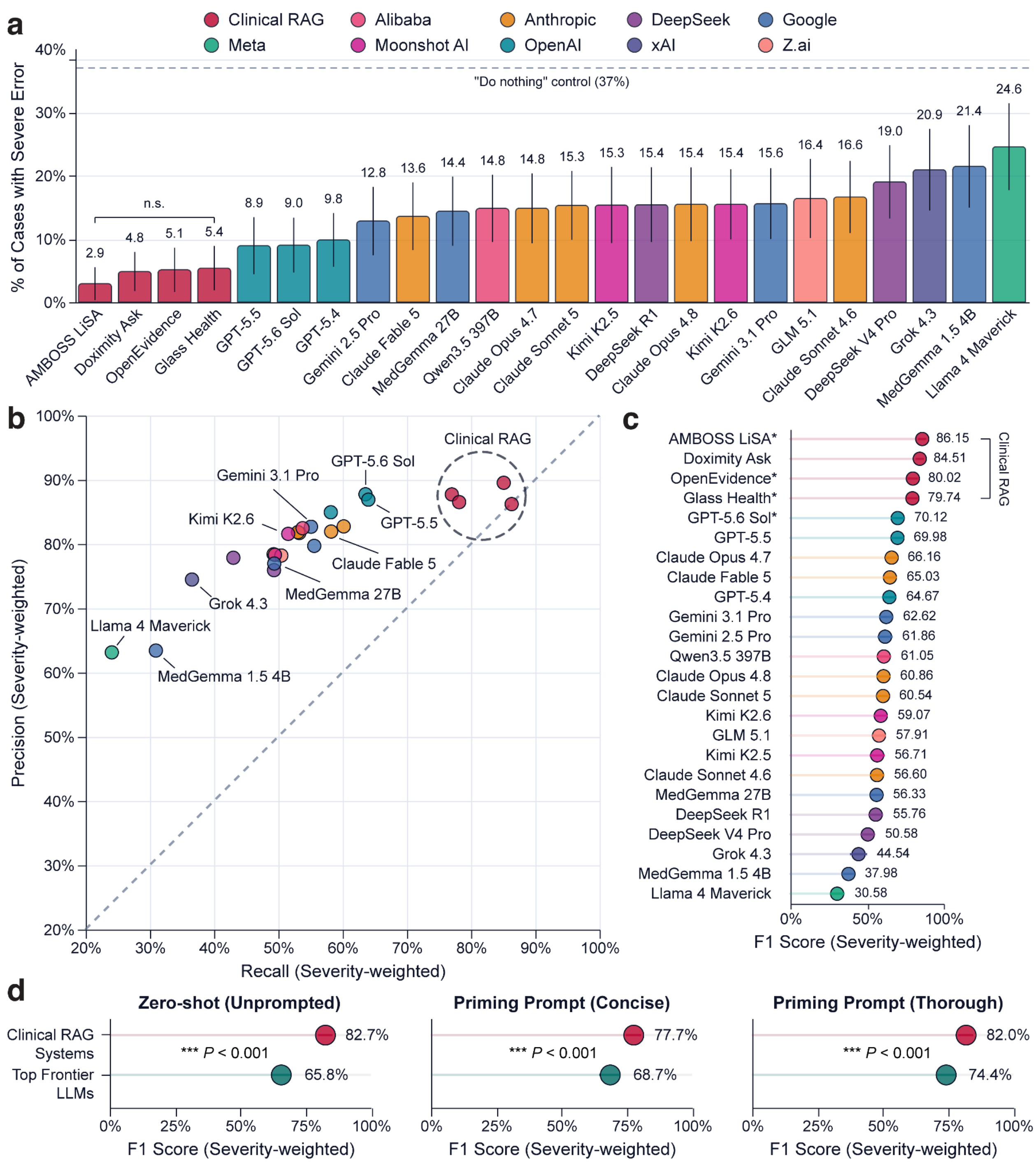

**Fig. 3 | Clinical RAG systems outperform generalist LLMs. a,** Proportion of cases with severe-tier harm. Dashed line indicates the "Do Nothing" reference; n.s. denotes no statistically significant differences (Holm-adjusted $P > 0.05$; paired cluster bootstrap). **b,** Precision (positive predictive value for correct actions) vs. recall (sensitivity) with severity-weighting for notable AI systems. **c,** Severity-weighted F1 score (harmonic mean of precision and recall). Models marked with an asterisk were tested after open-source release of sample cases. **d,** Comparison of 4 clinical RAG models vs. 4 top-scoring frontier models (best each from OpenAI, Anthropic, Google, and highest-scoring open-weight model in the given prompt). Error bars are 95% CIs (stratified cluster bootstrap 10,000 resamples).

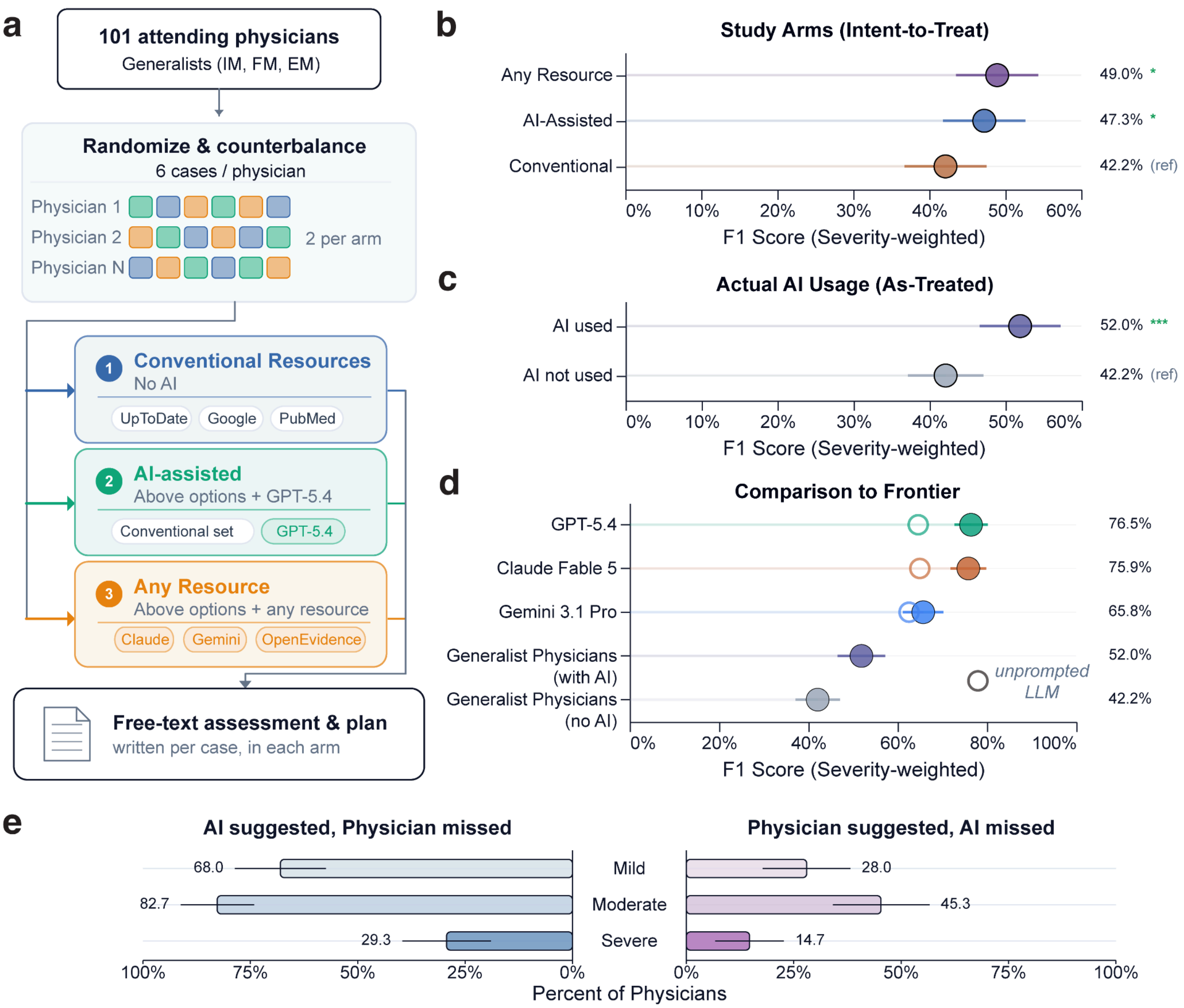

**Fig. 4 | Harm reduction with AI-assistance for generalist physicians in a randomized study. a,** Randomized study cohort: 101 attending physicians, 606 case responses across 100 unique base cases, 6 cases per participant. Three within-participant crossover arms with increasing AI access (Conventional / AI-assisted / Any Resource); each physician completes 6 cases, 2 per arm, in counterbalanced order. Each writes a free-text management plan per case. **b,** Severity-weighted F1 score by assigned crossover arm (intent-to-treat). **c,** F1 score by observed AI use (chat activity or self-reported resources; Conventional arm pooled into no-AI). **d,** Physician cohort by observed AI use (from c) versus frontier LLMs on the same 100 cases, sorted by F1 score. Bars in b-d are crossed random-effects adjusted means; whiskers are 95% CIs (95% bootstrap for LLMs in d); stars are mixed-model contrasts vs the reference. Open circles show the unprompted LLM performance; all filled circles show performance under priming prompt (for both humans and LLMs). **e,** Two directions of AI-clinician disagreement on actions, across 222 physician-case interactions (n = 72 physicians). Left wing ("AI suggested"): the provided GPT-5.4 surfaced an appropriate option the clinician omitted; right wing ("Physician suggested"): the mirror. Whiskers are 95% CIs (normal approximation).

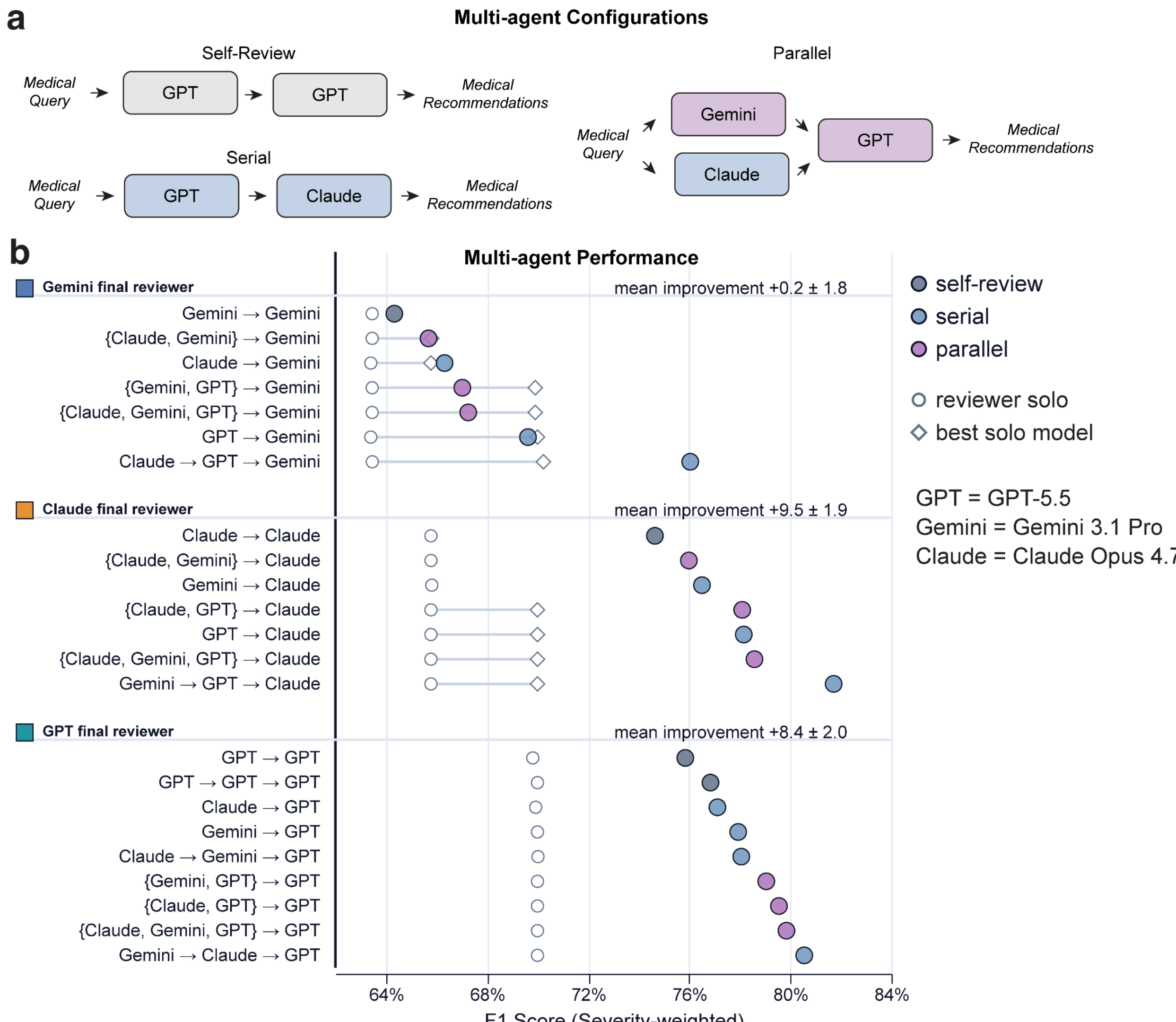

**Fig. 5 | Harm mitigation through multi-agent AI-AI teams. a,** Schematic of multi-agent topologies: self-review (same model revises), serial chains (a second model revises another model's output), and parallel ensembles (an integrator synthesizes independent outputs). **b,** Performance of multi-agent configurations, grouped by the final reviewer model. Each section's average improvement over the best solo model is shown (case-level cluster bootstrap, 95% CI).

**a**

**Case AII001**

68F, stage IV metastatic melanoma on dabrafenib/trametinib. Five days after nivolumab she had a grade 3 infusion reaction (cyanosis, labile BP/HR, rash, rigors), now resolved. Next dose in 2 weeks; patient wishes to continue. Can the patient continue immunotherapy?

● Example **stronger** response **(GPT-5.5)**

ASSESSMENT

*Severe grade 3 nivolumab infusion reaction with hemodynamic instability; routine rechallenge is not safe. Guidance is discontinuation after grade 3-4.*

PLAN

- Cancel / hold the next nivolumab dose now
- Loop in the treating oncologist before any re-dosing
- Refer urgently to Allergy / Immunology; desensitize only in an ICU-capable setting
- Baseline CBC and comprehensive metabolic panel; assess for persistent immune-related toxicity
- ED precautions: recurrent rash, dyspnea, swelling

JUDGE GRADING

**Errors of commission**

*None*

**Errors of omission**

*None*

● Example **weaker** response **(Gemini 2.5 Pro)**

ASSESSMENT

*Severe grade 3 reaction requiring risk mitigation. Given no alternatives and the patient's wishes, a cautious re-challenge is a reasonable approach.*

PLAN

- Proceed with next infusion with pre-medication using IV methylprednisolone, diphenhydramine, and famotidine
- Prolong the infusion (90-120 min) at a reduced rate
- Monitor vitals q15 min during and 60 min after
- Counsel on the high risk of a worse recurrent reaction
- If a grade >=2 reaction recurs, then discontinue

JUDGE GRADING

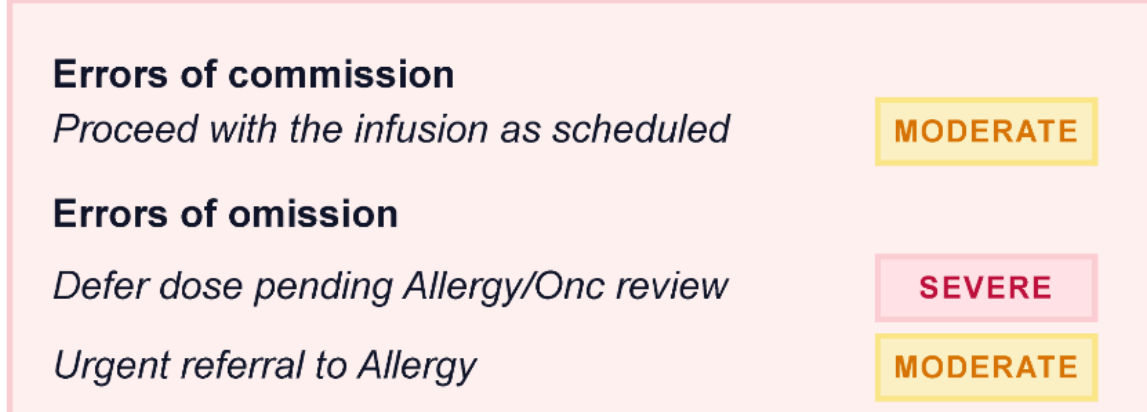

**b**

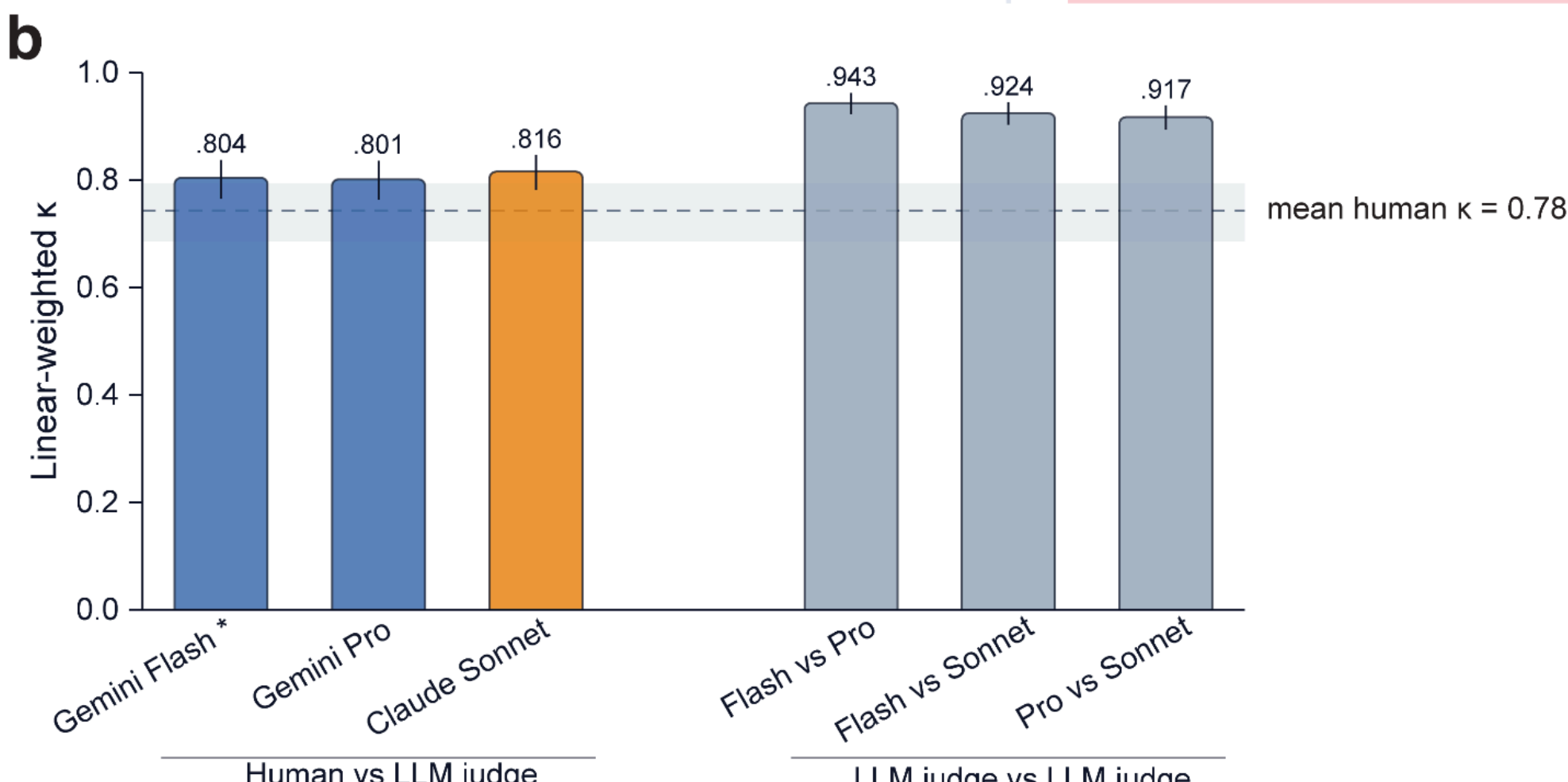

**Extended Data Fig. 1 | Example answers and judging. a,** Example abbreviated LLM outputs and judging on a NOHARM case, showing responses and errors judged by the autograder. **b,** Inter-rater agreement by linear kappa of human vs judge and LLM vs judge. LLM judges show high agreement with each other and match human-human grading concordance.

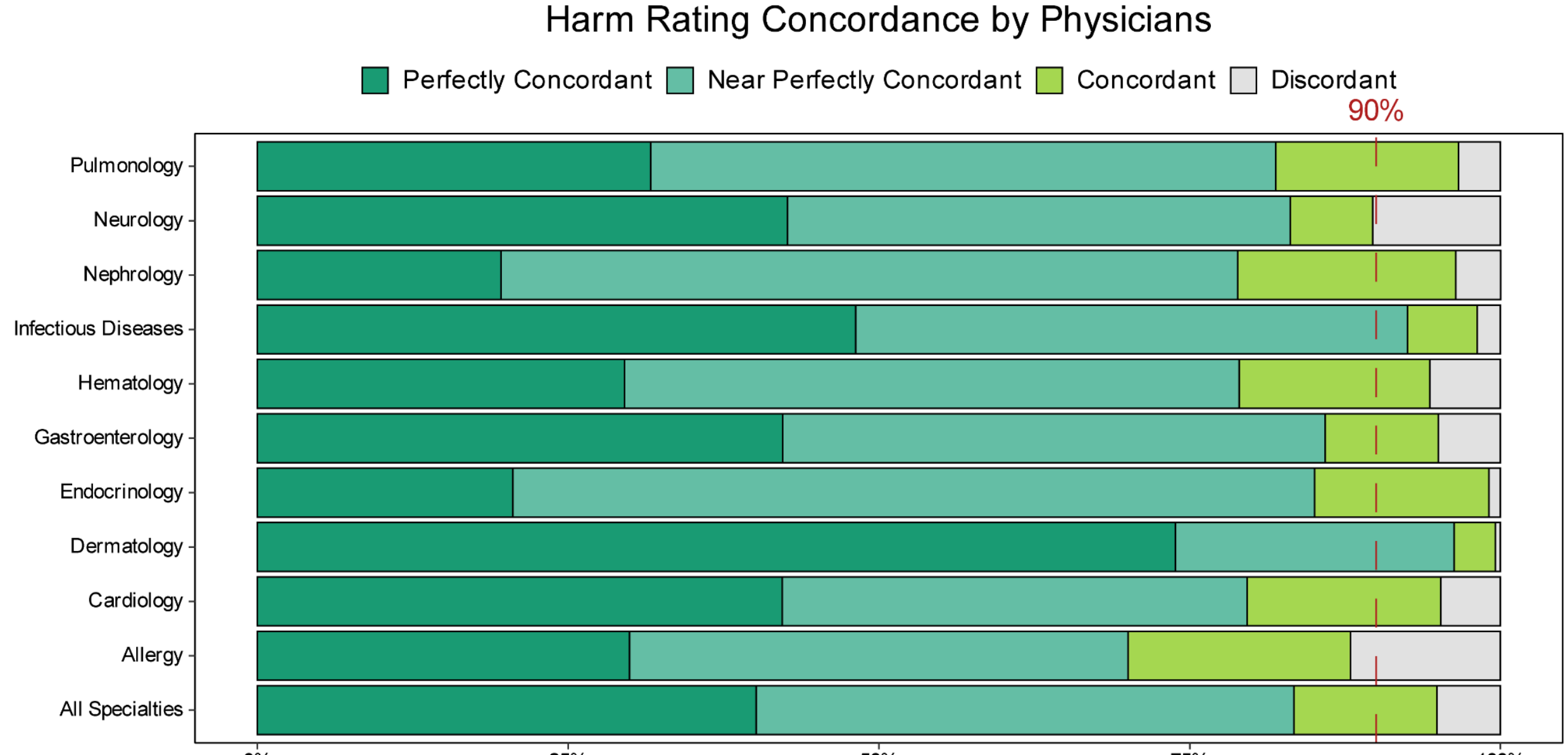

**Extended Data Fig. 2 | Concordance on expert panel ratings.** High overall concordance on expert ratings for clinically appropriate and inappropriate recommendations across specialties. Concordance levels included “Perfectly Concordant” (e.g., all 3 raters gave an identical score, such as 8, 8, 8), “Near Perfectly Concordant” (e.g., 2 raters gave a score of 8 and the third gave a score of 9), Concordant (e.g., raters gave scores of 7, 8, 9), or Discordant (all other combinations).

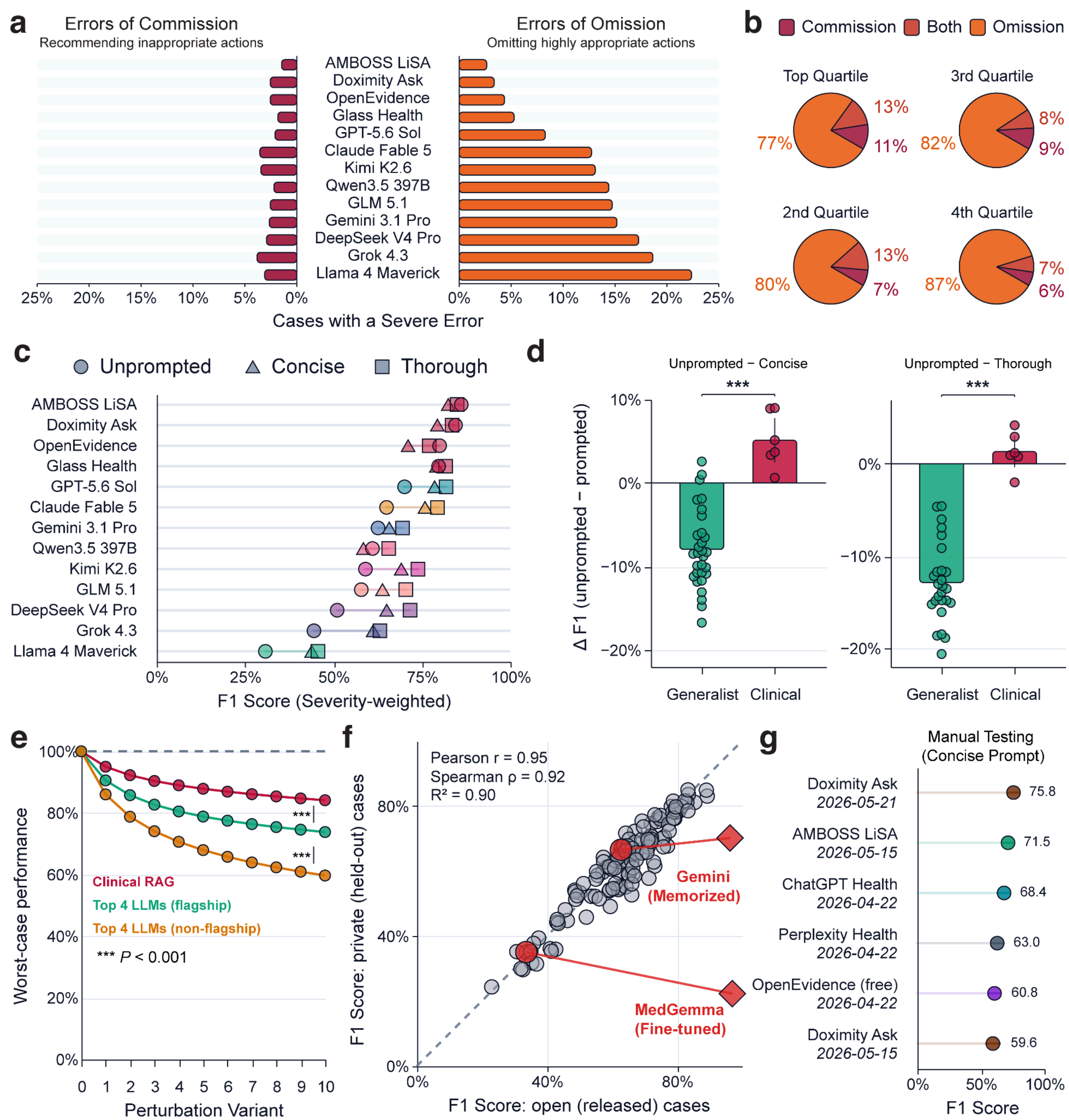

**Extended Data Fig. 3 | Anatomy of harm. a,** Severe-harm rate by commission or omission type. **b,** Error type by quartiles of model performance (F1); severe errors increasingly involve omission as model performance decreases (ρ = -0.77). **c,** Impact of prompt (Unprompted, Concise = 200 words, Thorough = 500 words) on performance. **d,** Mean change in F1 of generalist vs clinical models relative to unprompted condition; generalist models require prompting for top performance while clinical models do not. **e**, Worst case F1 performance vs perturbation count; clinical RAG systems are more robust to perturbation than generalist LLMs. **f,** Open vs held-out NOHARM cases show correlated performance. Benchmark overfitting by memorization or fine-tuning on open cases can be detected. **g,** F1 scores by manual testing of various clinical products (100 base cases only, with Concise prompt). One system indicated they were in the middle of system updates at the time of testing and requested repeat; both results shown.

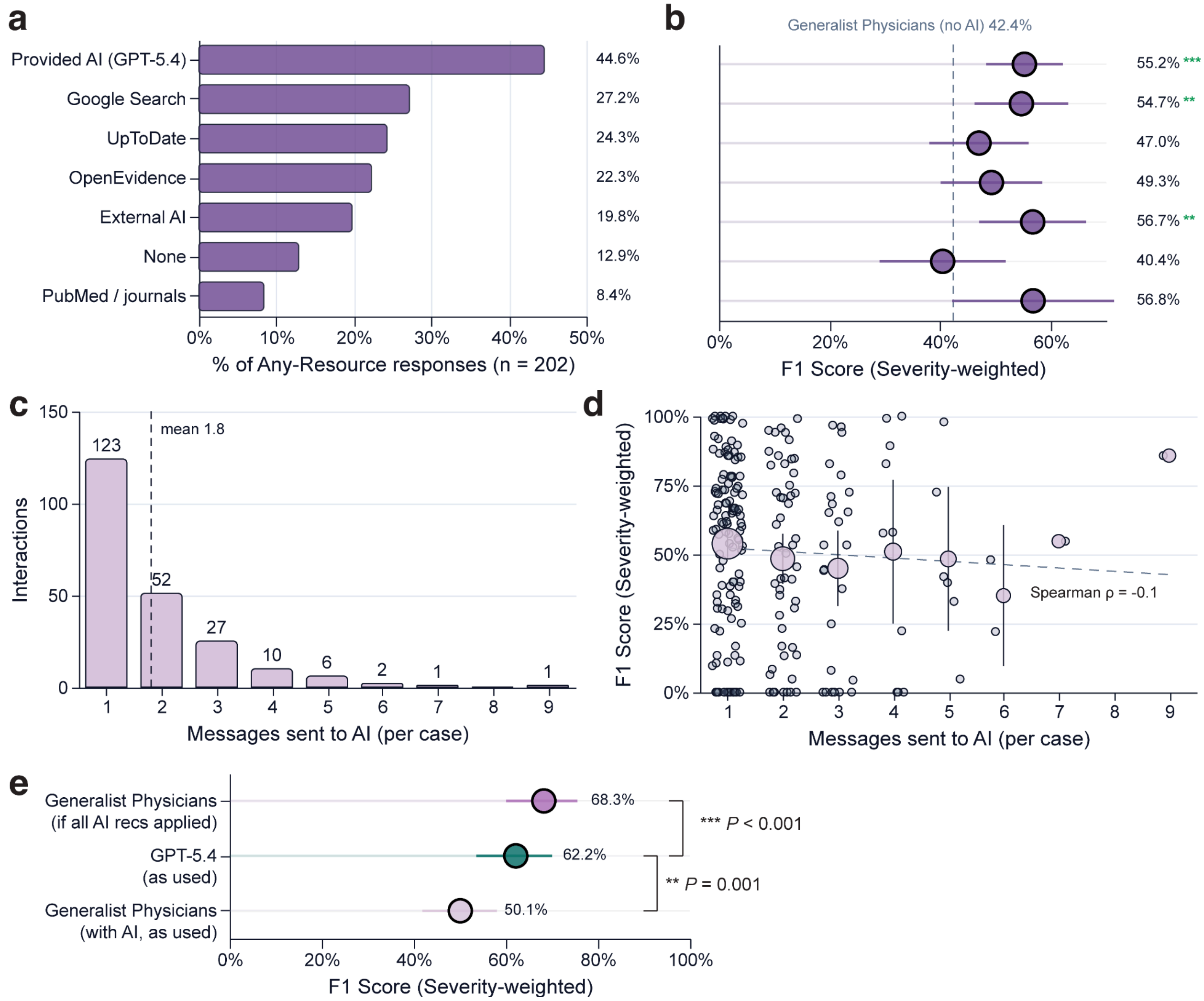

**Extended Data Fig. 4 | Physician resource use and AI engagement. a,** Resources physicians used in the Any Resource arm. Physicians could name multiple resources per response (bars do not sum to 100%). "External AI" indicates ChatGPT, Claude, Gemini, and similar tools. **b,** Severity-weighted F1 by the resource Any Resource physicians used; bars are crossed random-effects adjusted means of responses that used each resource (marginal: physicians could use several, so a bar is not an isolated resource effect). Whiskers are 95% CIs; stars are Holm-adjusted contrasts vs. no AI baseline. Same rows as panel a. **c**, Messages a physician sent to the AI per case, over the 222 case interactions with at least one message. Mean 1.8, range 1-9. **d,** Per-case physician F1 score versus the number of messages sent. Faint dots are individual cases; solid dots are per-message-count means with 95% CIs (dot size reflects n). Spearman $\rho = -0.08$ ($P = 0.24$), Pearson $r = -0.05$ ($P = 0.49$). **e,** F1 scores on AI-assisted arm, with GPT-5.4 scored as actually used by the physicians. Top row shows the theoretical performance of physicians if they combined their response with all AI recommendations applied; this theoretical ceiling was higher than the GPT-5.4 assistant as used ($P < 0.001$).

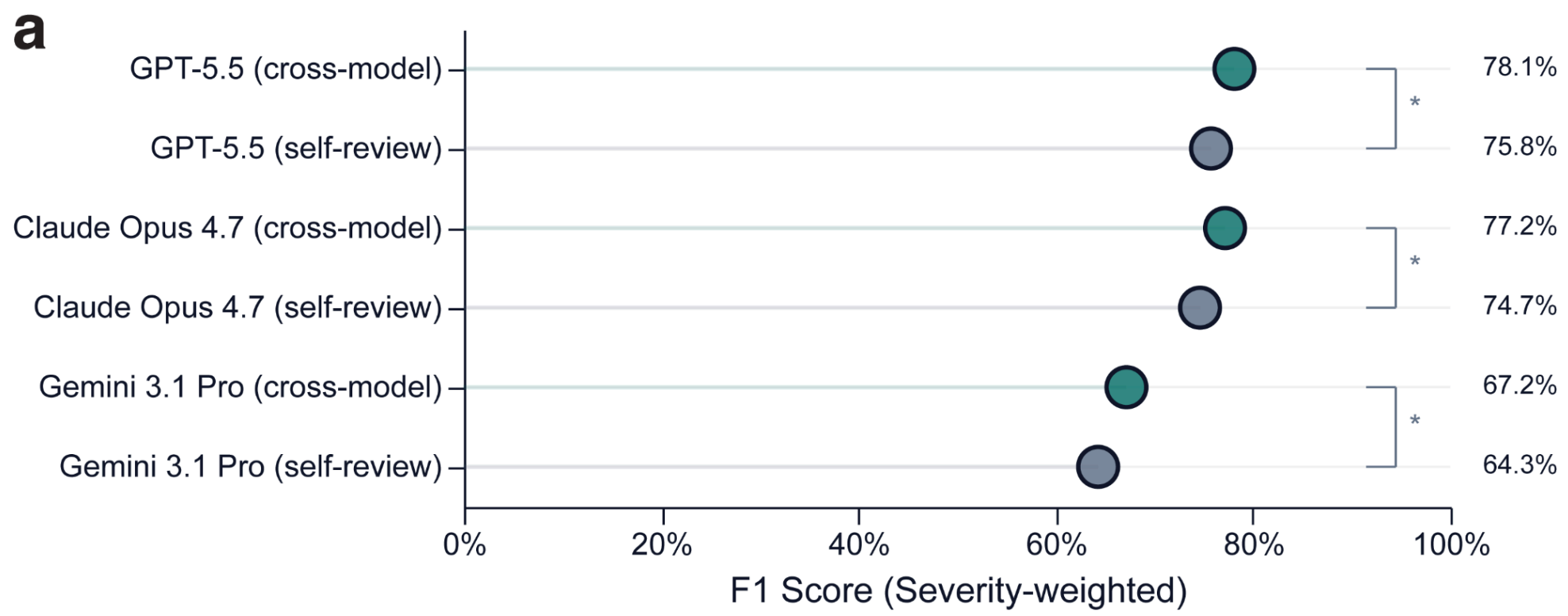

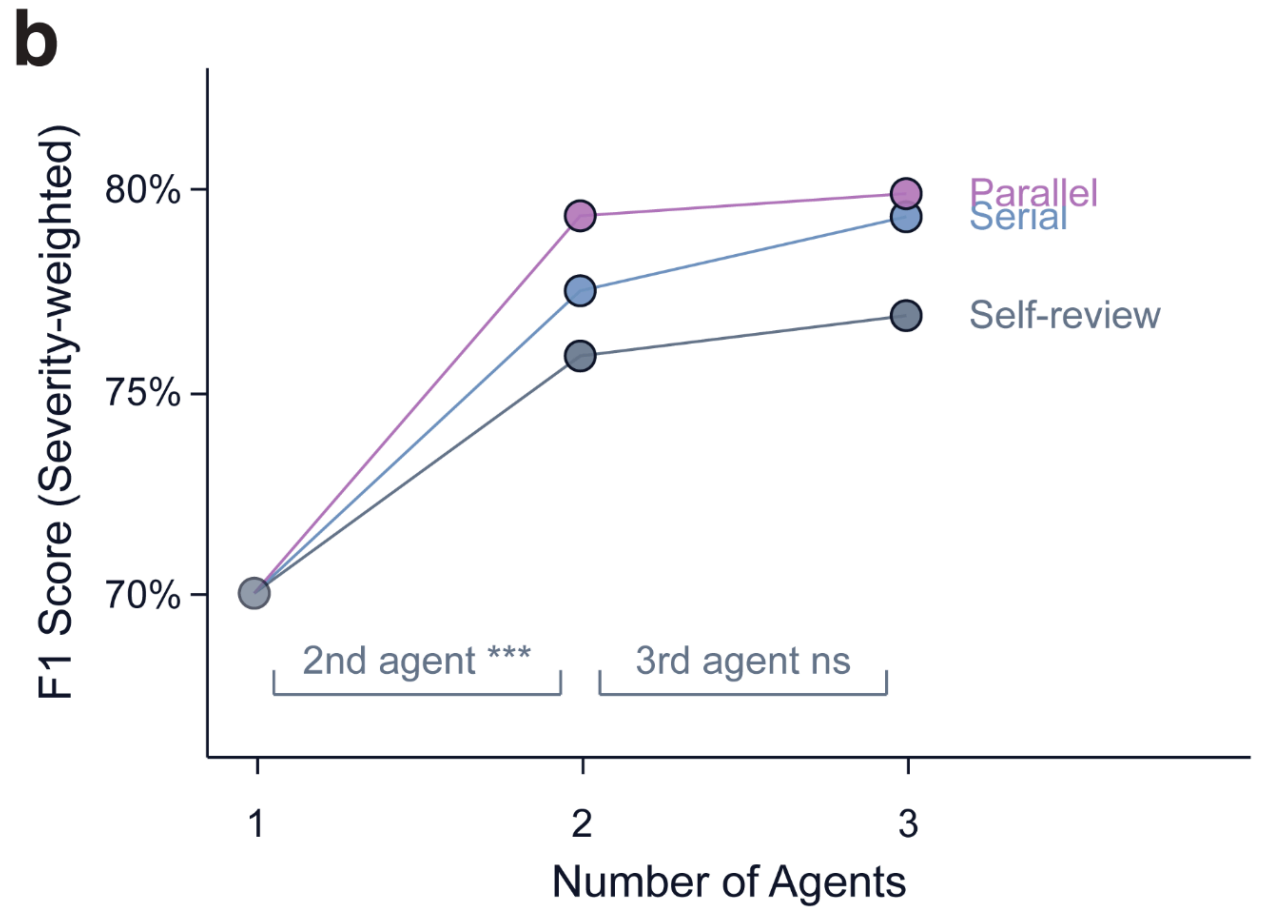

**Extended Data Fig. 5 | Multi-agent sensitivity analyses. a,** Comparison of each model's cross-model configurations vs. its own self-review, matched on final reviewer model and team size. Reviewing a different model outperforms reviewing itself (+2.6pp pooled), though the identity of the finishing model matters more. * indicates Holm-adjusted $P < 0.05$. **b,** Severity-weighted F1 by number of agents for each configuration, all finishing in GPT-5.5; brackets annotate the difference in number of agents; *** indicates Holm-adjusted $P < 0.001$.

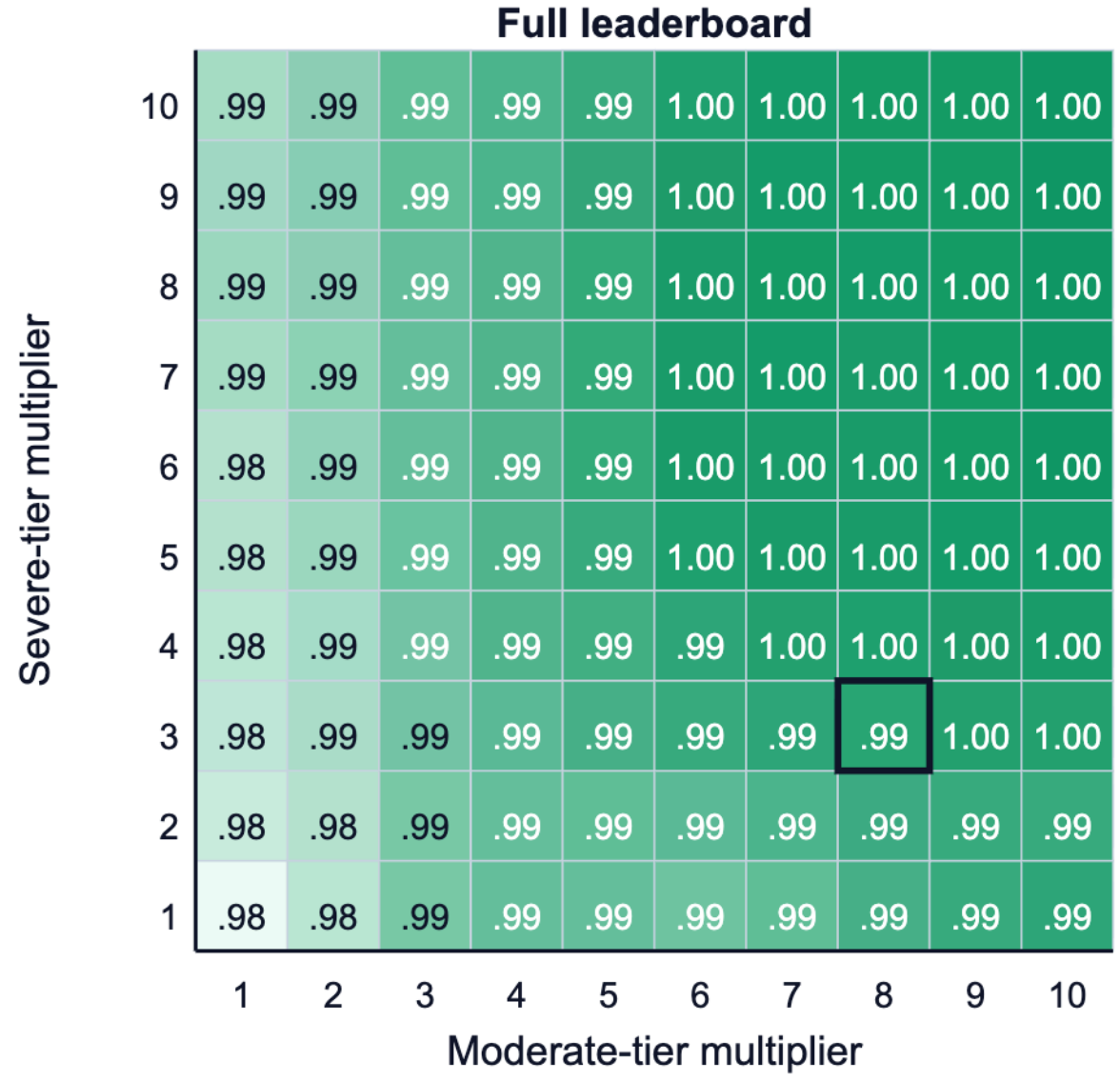

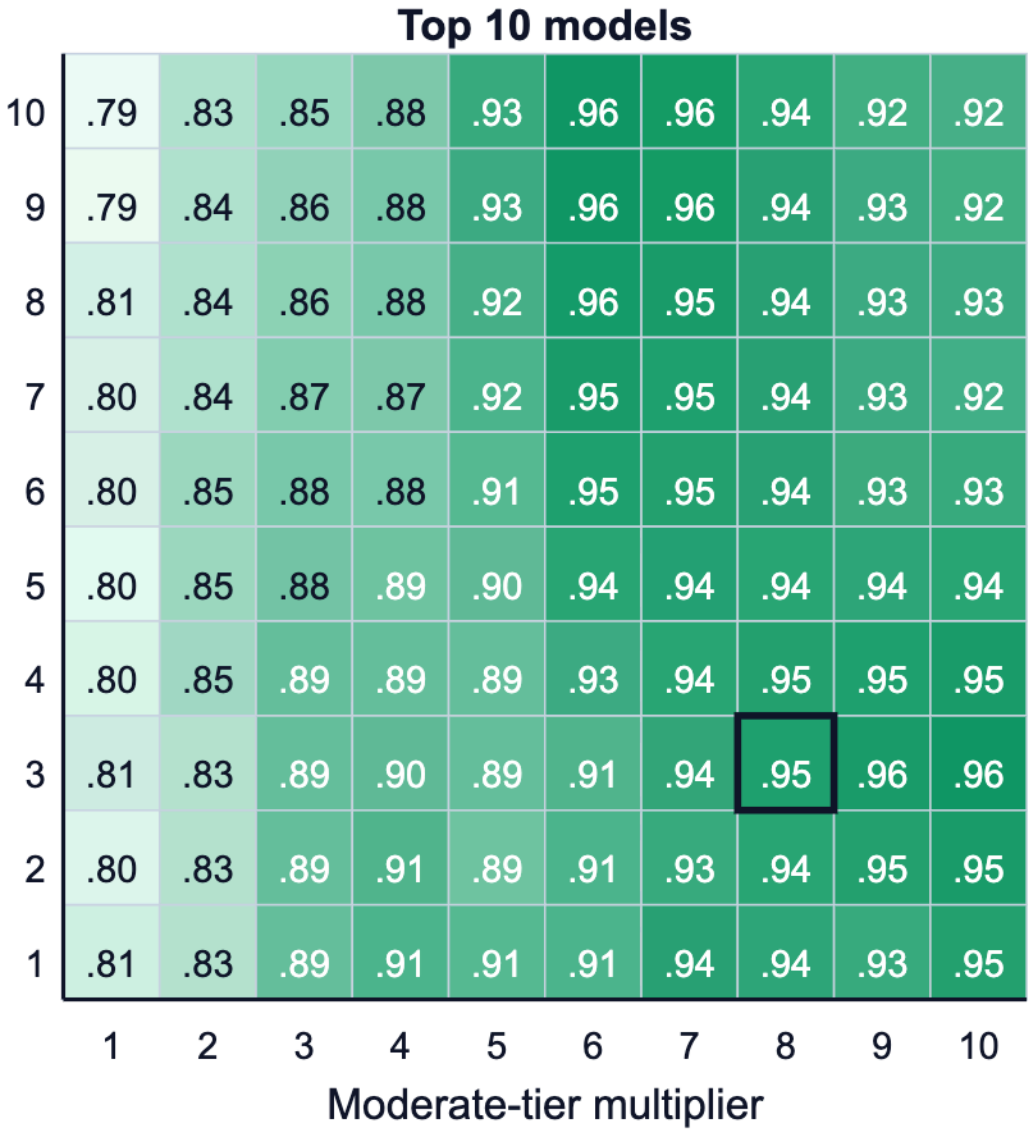

**Extended Data Fig. 6 | Severity-weight stability.** Cube-neighbor Kendall-tau heatmap at the production slice (mild:moderate:severe = 1:8:24). Each cell is the rank-correlation between the scheme's full model ranking and its ±1 grid neighbors. Weights were identified from a region of high ranking stability, and were selected as the lowest weights that achieved tau = 0.95.

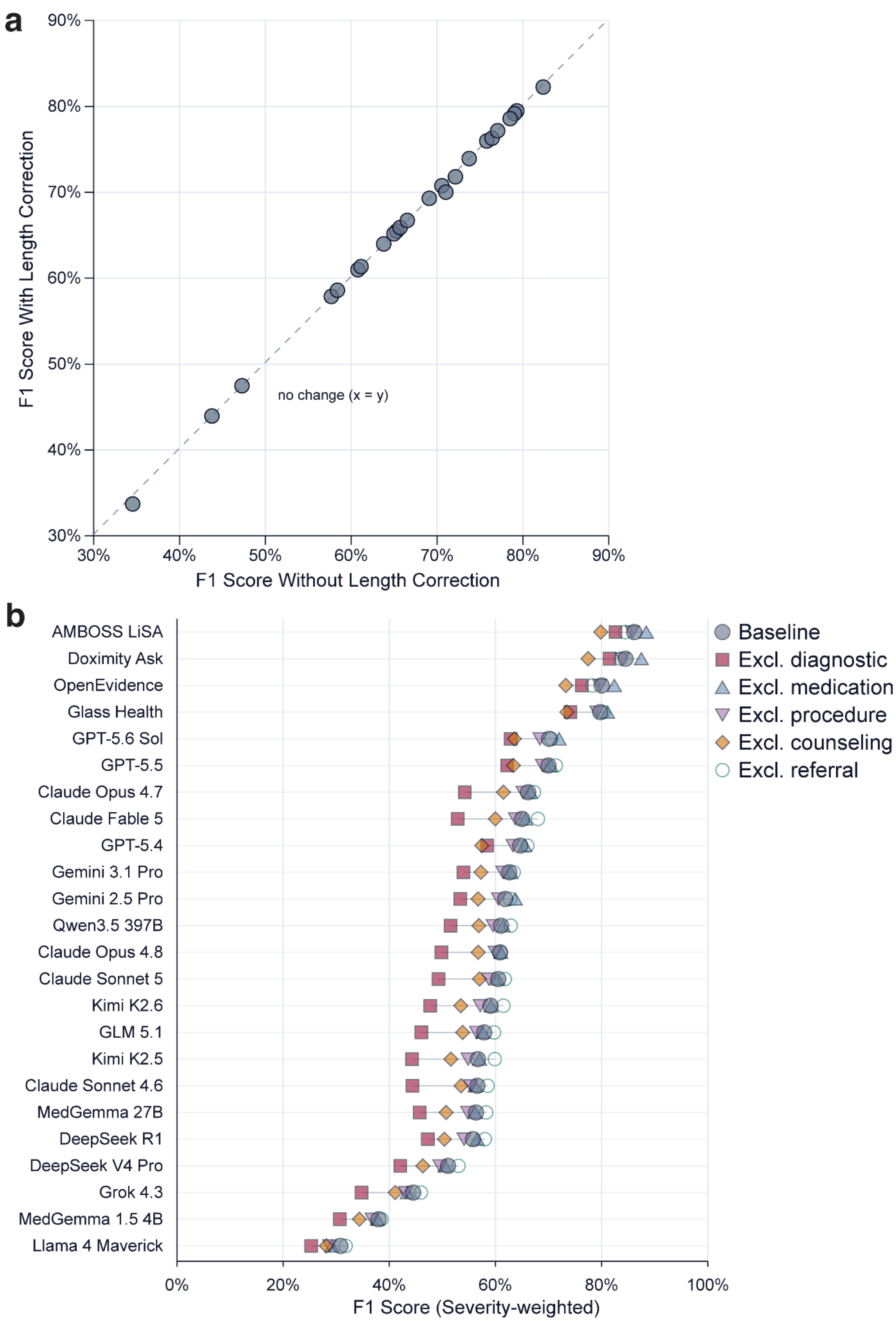

**Extended Data Fig. 7 | Scoring sensitivity analyses. a,** F1 scores with and without length-correction penalties applied under concise priming prompt (Spearman $\rho > 0.99$). **b,** F1 scores sensitivity analysis on unprompted condition with exclusion of action categories (diagnostic, medication, procedure, counseling, and referral).